\documentclass[sigconf, nonacm]{acmart}
\graphicspath{{./figures/}}

\usepackage{amsfonts}
\usepackage{algorithmic}
\usepackage{caption}
\usepackage{subcaption}

\usepackage[algo2e,ruled,vlined]{algorithm2e}
\SetCommentSty{scriptsize}
\SetKwComment{tcc}{\{}{\}}
\SetKwIF{If}{ElseIf}{Else}{if}{}{else if}{else}{endif}
\SetKwFor{While}{while}{}{endw}
\SetKwFor{ForEach}{for each}{}{endfch}

\newtheorem{exmp}{Example}[section]

\newcommand*\Laplace{\mathop{}\!\mathbin\bigtriangleup}
\newcommand*\Div{\mathop{}\!\mathbin\bigtriangledown\cdot}
\newcommand*\Grad{\mathop{}\!\mathbin\bigtriangledown}

\begin{document}

\title{Reducing Communication in Algebraic Multigrid with Multi-step Node Aware Communication}

\author{Amanda Bienz}
\email{bienz2@illinois.edu}
\author{William D. Gropp}
\email{wgropp@illinois.edu}
\author{Luke N. Olson}
\email{lukeo@illinois.edu}
\affiliation{\department{Department of Computer Science}
  \institution{University of Illinois at Urbana-Champaign}
  \city{Urbana}
  \state{Illinois}
  \postcode{61801}
}

\renewcommand{\shortauthors}{Bienz, Gropp, and Olson}

\begin{abstract}
Algebraic multigrid (AMG) is often viewed as a scalable $\mathcal{O}(n)$ solver
for sparse linear systems.  Yet, parallel AMG lacks scalability due to
increasingly large costs associated with communication, both in the initial
construction of a multigrid hierarchy as well as the iterative solve phase.  This work introduces a
parallel implementation of AMG to reduce the cost of
communication, yielding an increase in scalability.

Standard inter-process communication consists of sending data regardless of the
send and receive process locations.  Performance tests show
notable differences in the cost of intra- and inter-node communication, motivating a
restructuring of communication.  In this case, the communication schedule takes advantage of the less costly
intra-node communication, reducing both the number and size of inter-node
messages.  Node-centric communication extends to the range of components in
both the setup and solve phase of AMG, yielding an increase in the weak and
strong scalability of the entire method.
\end{abstract}



\maketitle

\section{Introduction}

Algebraic multigrid (AMG)~\cite{McRu1982, BrMcRu1984, RuStu1987} is an iterative
solver for sparse linear systems, such as those arising from discretized partial
differential equations.  AMG targets linear cost in the number of unknowns and
is dominated in cost by sparse matrix operations such as the sparse
matrix-matrix (SpGEMM) multiply and sparse matrix-vector (SpMV) multiplication.
As state-of-the-art supercomputers are continuously increasing in performance
capabilities, there is pressure on numerical methods to fully exploit the
potential of these machines.  Due to large costs associated with parallel
communication, for example on the coarse levels~\cite{AMGHPC}, AMG lacks
parallel scalability.  This paper explores a method of altering the parallel
implementation of communication throughout AMG to improve both performance and
scalability.

\begin{algorithm2e*}[ht!]
  \DontPrintSemicolon  \KwIn{$A$\tcc*{sparse system matrix}}
  \BlankLine	\KwOut{    \begin{tabular}[t]{l}
      $A_{0}$, $A_{1}$, \ldots, $A_{N}$\\
      $P_{0}$, $P_{1}$, \ldots, $P_{N-1}$
    \end{tabular}
    \tcc*{hierarchy of sparse operators}
  }
  \BlankLine  $A_{0} \leftarrow A$\;
  $\ell \leftarrow 0$\;
  \While {$\left|A_{\ell}\right| > \textnormal{max\_coarse}$} {
      $S_{\ell} \leftarrow \texttt{strength}(A_{\ell})$
            \tcc*{strength-of-connection}
      \If{Aggregation}{
        $Agg_{\ell} \leftarrow \texttt{splitting}(S_{\ell})$\tcc*{partition nodes}
        $P_{\ell} \leftarrow \texttt{interpolation}(Agg_{\ell})$\tcc*{form interpolation}
      }
      \Else{
        $C_{\ell}, F_{\ell} \leftarrow \texttt{splitting}(S_{\ell})$\tcc*{partition nodes}
        $P_{\ell} \leftarrow \texttt{interpolation}(C_{\ell}, F_{\ell})$\tcc*{form interpolation}
      }
      $A_{\ell + 1} \leftarrow P_{\ell}^{T} \cdot A_{\ell} \cdot P_{\ell}$
            \tcc*{coarse operator, Galerkin product}
      $\ell \leftarrow \ell + 1$\;
  }
    \caption{AMG Setup: \texttt{setup}}\label{alg:setup}
\end{algorithm2e*}

\begin{figure}[h]
    \centering
    \includegraphics[width=0.375\textwidth]{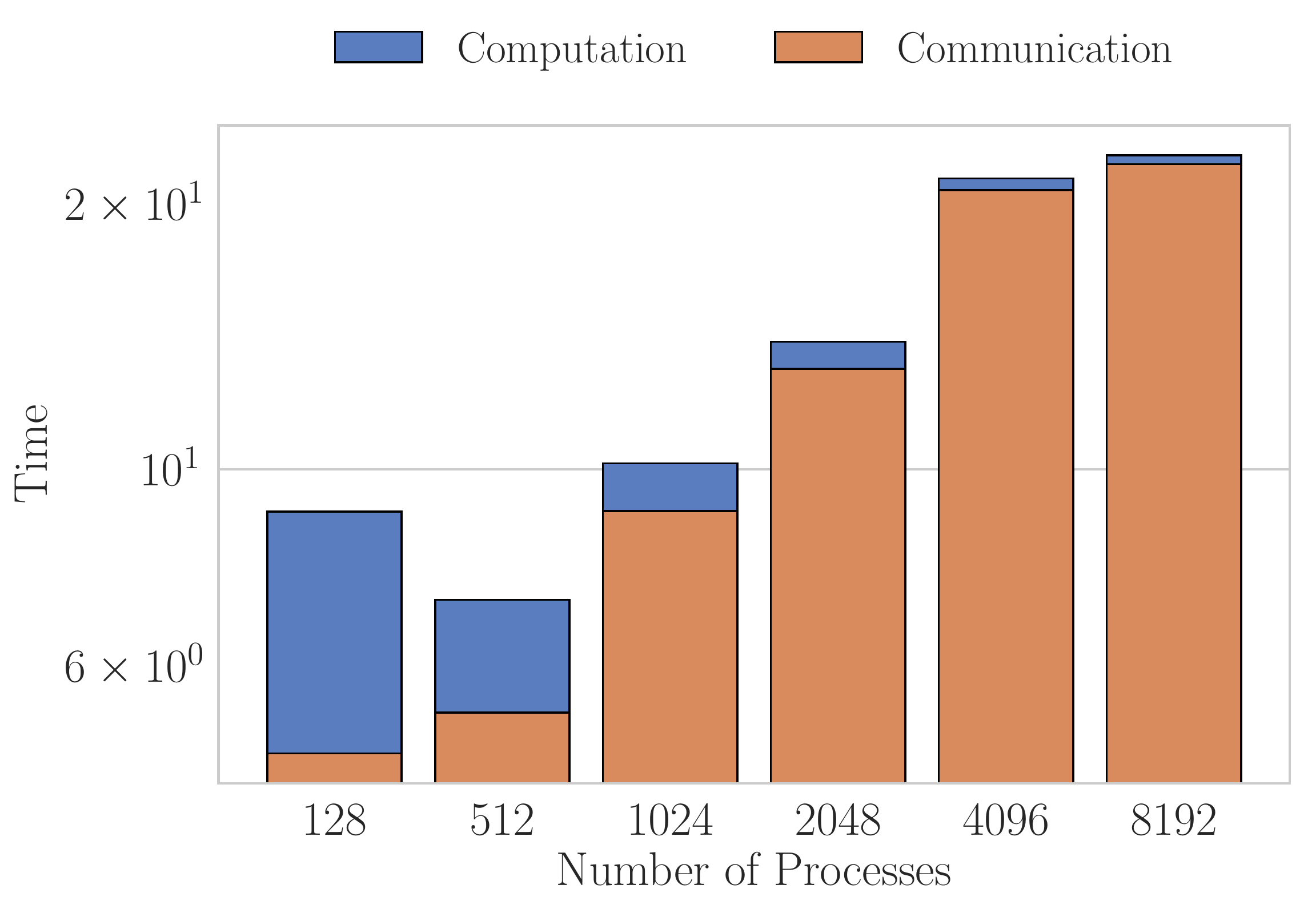}
    \caption{The total time required to solve a three-dimensional Laplacian
    system described in Example~\ref{exmp:laplace}}\label{figure:amg_scale}
    \Description{3D Laplacian Example.}
\end{figure}

Standard parallel algebraic multigrid typically exhibits strong scaling to five
or ten thousand degrees-of-freedom per core before communication costs outweigh
local computation.  Further extending the core-count yields an increase in total
solve time due to dominant communication costs.  Figure~\ref{figure:amg_scale}
shows the time required to solve a Laplacian system, created with MFEM, and
described in detail in Example~\ref{exmp:laplace}. The timings are partitioned
into local computation and inter-process communication costs.  As the processor
count reaches the strong scaling limit, communication becomes increasingly
dominant.  The problem scales to $512$ processes, after which communication
costs outweigh any reductions in local computation.

Most methods for reducing communication costs in AMG focus on a redesign of the
method or on the underlying sparse matrix operations.  Aggressive coarsening,
for example, reduces the dimensions of coarse levels at a faster rate, yielding
reduced density and communication
requirements~\cite{AggCoarse,AggCoarse2,DistTwo}.  Similarly, the smoothed
aggregation solver allows large aggregates, coarsening a larger number of fine
points into a single coarse point~\cite{1592718,NonGal_Treister}.  Small
non-zeros resulting from fill-in on coarse levels may be systematically removed,
adding sparsity into coarse-grid
operators~\cite{NonGal_Schroder,NonGal_Treister,Bienz_sparsegal}.  Furthermore,
matrix ordering and graph partitioning yield reduced communication costs
throughout sparse matrix
operations~\cite{Pinar,Hendrickson,BisselingDataDist,TwoDimPart,HypergraphPart},
and coarse level repartitioning has potential to reduce the cost of the solver.
Likewise, coarse level redistribution and duplication of the coarsest level
solves yield large reductions in communication time.  The approach presented in
this paper augments these approaches, reducing off-node message counts and sizes
through aggregation of data.

Topology-aware methods and message agglomeration are commonly used to reduce
communication costs in MPI applications.  Topology-aware task mapping minimizes
message hop counts, reducing the cost associated with
communication~\cite{TopoTaskMap,SubcommTaskMap}.  Message agglomeration is
commonly used to reduce the cost of communication, for example in MPI
collectives~\cite{Solomonik,Kielmann,Karonis,Sack}.  The Tram
library~\cite{Tram} explores agglomeration of point-to-point messages, by
streamlining messages between neighboring processes~\cite{Tram}.

This paper presents, analyzes, and evaluates a method for reducing
communication costs in both the setup and solve phases of parallel algebraic
multigrid through agglomeration of messages among nodes.
Section~\ref{section:background} covers algebraic multigrid and common parallel
implementations.  Section~\ref{section:node_aware} focuses on the node-aware
communication algorithm, with two variations described in
Sections~\ref{section:three_step} and~\ref{section:two_step}.
Section~\ref{section:perf} presents performance models that differentiate
between communication strategies.  Section~\ref{section:results} covers
numerical experiments in support of the approach, and
Section~\ref{section:conclusion} contains concluding remarks and future
directions.

\section{Background}\label{section:background}

\begin{algorithm2e*}[!t]
  \DontPrintSemicolon	\KwIn{$A_{\ell}$, $P_{\ell}$, $x_{\ell}$, $b_{\ell}$
        \tcc*{system, interpolation, initial solution, right-hand side at level $\ell$}
  }
  \BlankLine	\KwOut{$x_{\ell}$ \tcc*{updated solution vector}\\
  }
  \BlankLine  \If {$\ell = N$}{
      $\texttt{solve}$ $A_{\ell}x_{\ell} = b_{\ell}$\;
  }
  \Else{
      $\texttt{relax}$ $A_{\ell}x_{\ell} = b_{\ell}$
        \tcc*{pre-relaxation}
      $r_{\ell} \leftarrow b_{\ell} - A_{\ell} \cdot x_{\ell}$
        \tcc*{calculate residual}
      $r_{\ell+1} \leftarrow P^{T}_{\ell} \cdot r_{\ell}$
        \tcc*{restrict residual}
      $e_{\ell+1} \leftarrow \texttt{solve}(A_{\ell+1}, P_{\ell+1}, 0, r_{\ell+1})$
        \tcc*{coarse-grid solve}
      $e_{\ell} \leftarrow P_{\ell} \cdot e_{\ell+1}$
        \tcc*{interpolate error}
      $x_{\ell} \leftarrow x_{\ell} + e_{\ell}$
        \tcc*{update solution}
      $\texttt{relax}$ $A_{\ell}x_{\ell} = b_{\ell}$
        \tcc*{post-relaxation}
  }
  \caption{AMG Solve: \texttt{solve}}\label{alg:solve}
\end{algorithm2e*}

Throughout this paper, algebraic multigrid methods are analyzed with regards to
the three-dimensional Laplacian in Example~\ref{exmp:laplace}\@.  This system is
representative of the types of problems that are often solved with AMG.
\begin{exmp}\label{exmp:laplace}
    Let the system $Ax = b$ result from a finite element discretization of the
    Laplace problem $-\Laplace u = 1$, created with MFEM~\cite{mfem-library}.  The
    linear system is created with MFEM's escher-p3 mesh, a three-dimensional
    mesh consisting of unstructured elements with structured refinement.
    Furthermore, this system consists of $1\,884\,545$ degrees-of-freedom and
    $27\,870\,337$ non-zeros, unless otherwise specified.  The associated
    Ruge-St\"{u}ben hierarchies are created with HMIS coarsening and extended+i
    interpolation, while the smoothed aggregation solver forms aggregates based
    on a distance-2 maximal independent set (MIS-2) of the graph.  Both
    classical and smoothed aggregation hierarchies use a strength tolerance of
    $0.25$.  All timings are performed on $8\,192$ processes of Blue Waters~\cite{BlueWaters, bw-in-vetter13}, a
    Cray XK/XE supercomputer at the National Center for Supercomputing
    Applications, unless stated otherwise.
\end{exmp}

Algebraic multigrid consists of forming a hierarchy of successively coarser
levels, followed by an iterative a solve phase.  Common algorithms for
constructing an algebraic multigrid hierarchy include the Ruge-St\"{u}ben
solver~\cite{RuStu1987} and the smoothed aggregation solver~\cite{1592718}.  This setup phase,
described in Algorithm~\ref{alg:setup}, consists four methods: \texttt{strength},
\texttt{splitting}, \texttt{interpolation}, and $P^{T} \cdot A \cdot P$,
regardless of the solver used.  First, the \texttt{strength} function determines
nodes that are strongly connected to one another.  The resulting
strength-of-connection matrix is then partitioned in \texttt{splitting} to
determine which nodes influence each coarse degree-of-freedom.  The
\texttt{interpolation} function uses the partitioned nodes to form a transfer
operator, which projects data between the fine and coarse nodes.  Finally, the
coarse-grid operator is formed through the Galerkin product, $P^{T} \cdot A
\cdot P$.

The underlying algorithms for \texttt{splitting} and \texttt{interpolation}
are solver dependent.  The Ruge-St\"{u}ben solver partitions nodes into coarse
(C) and fine (F) points.  The interpolation operator projects C-points directly
between fine and coarse levels, while F-points influence neighboring coarse
nodes.  Alternatively, the smoothed aggregation solver partitions the nodes into
groups of aggregates, each corresponding to a single coarse degree-of-freedom.
The transfer operator is initially created as a point-wise constant, with a
single column holding each aggregate.  Near-nullspace candidates are then fit to
the operator, and finally, the resulting matrix is smoothed.

\begin{figure}[ht!]
    \centering
    \begin{subfigure}{0.49\textwidth}
        \centering
        \includegraphics[width=0.75\linewidth]{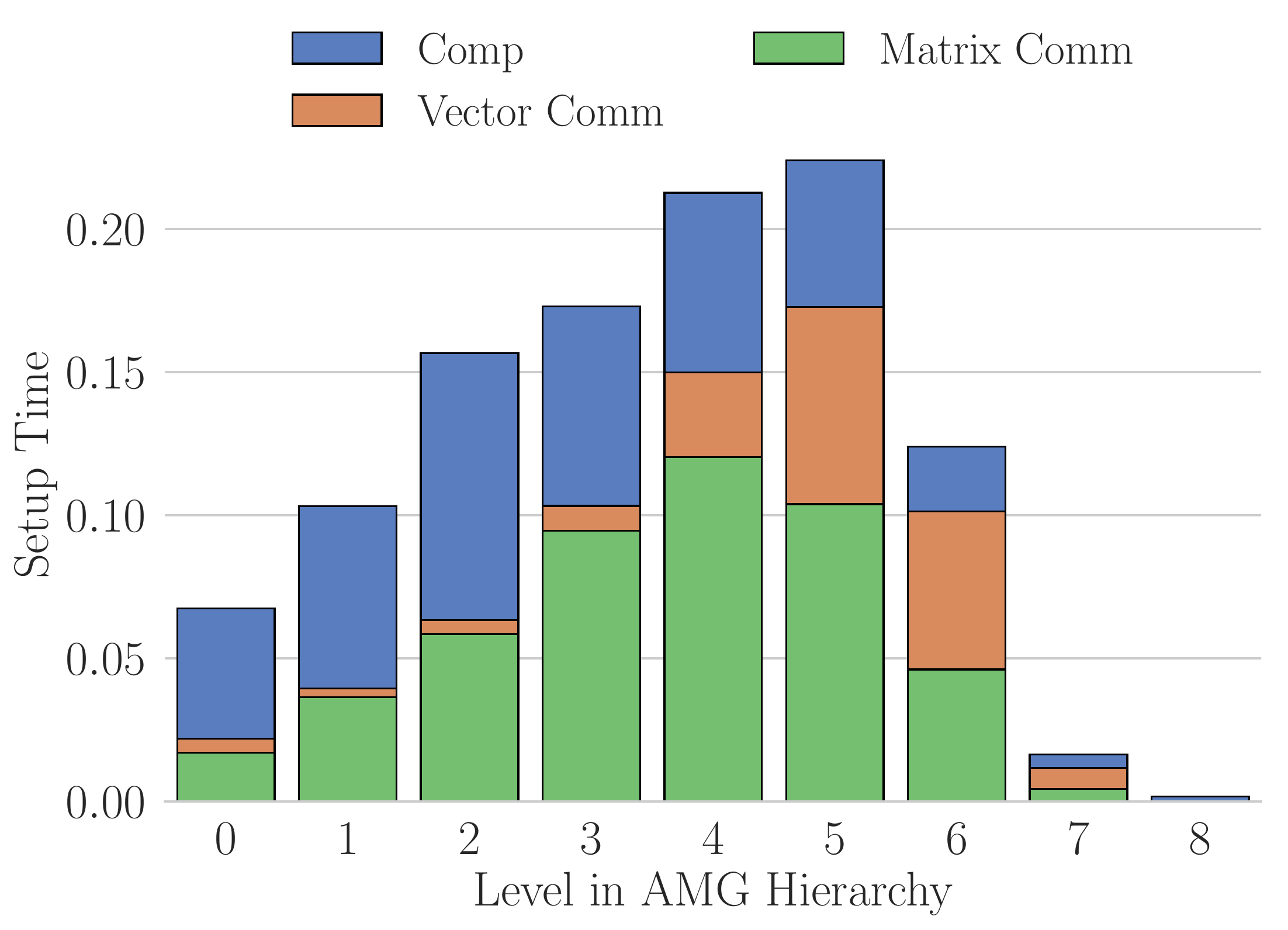}
        \caption{Ruge-Stuben}\label{figure:profile_rss_comm_setup}
    \end{subfigure}
        \begin{subfigure}{0.49\textwidth}
        \centering
        \includegraphics[width=0.75\linewidth]{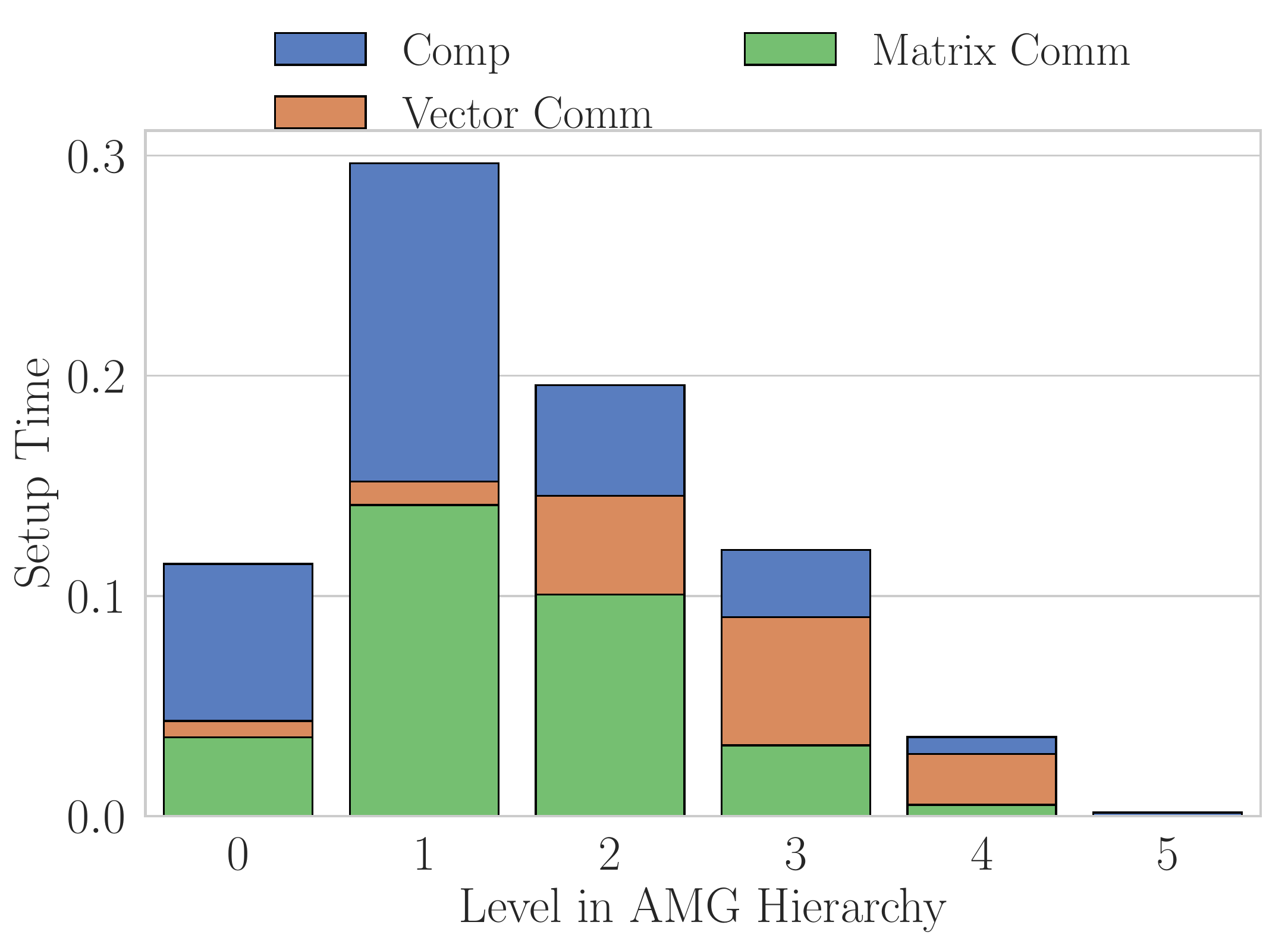}
        \caption{Smoothed Aggregation}\label{figure:profile_sas_comm_setup}
    \end{subfigure}
    \caption{The setup phase cost for various AMG hierarchies for the system from
    Example~\ref{exmp:laplace}.  The total cost is partitioned by level, and
    further split into communication and local computation.}\label{figure:setup_comm}
    \Description{AMG Setup Cost.}
\end{figure}

The construction of a parallel algebraic multigrid hierarchy requires both
local computation as well as point-to-point communication, specifically
communication of both vectors and sparse matrices.
Figure~\ref{figure:setup_comm} partitions the per-level cost of constructing a
hierarchy for Example~\ref{exmp:laplace}, vector
communication, and sparse matrix communication.  Irrespective of which setup
algorithm is used, the cost of hierarchy construction is split into local
computation and MPI communication, with communiction dominating coarse level
setup cost.

\begin{figure}[h]
    \centering
    \includegraphics[width=0.375\textwidth]{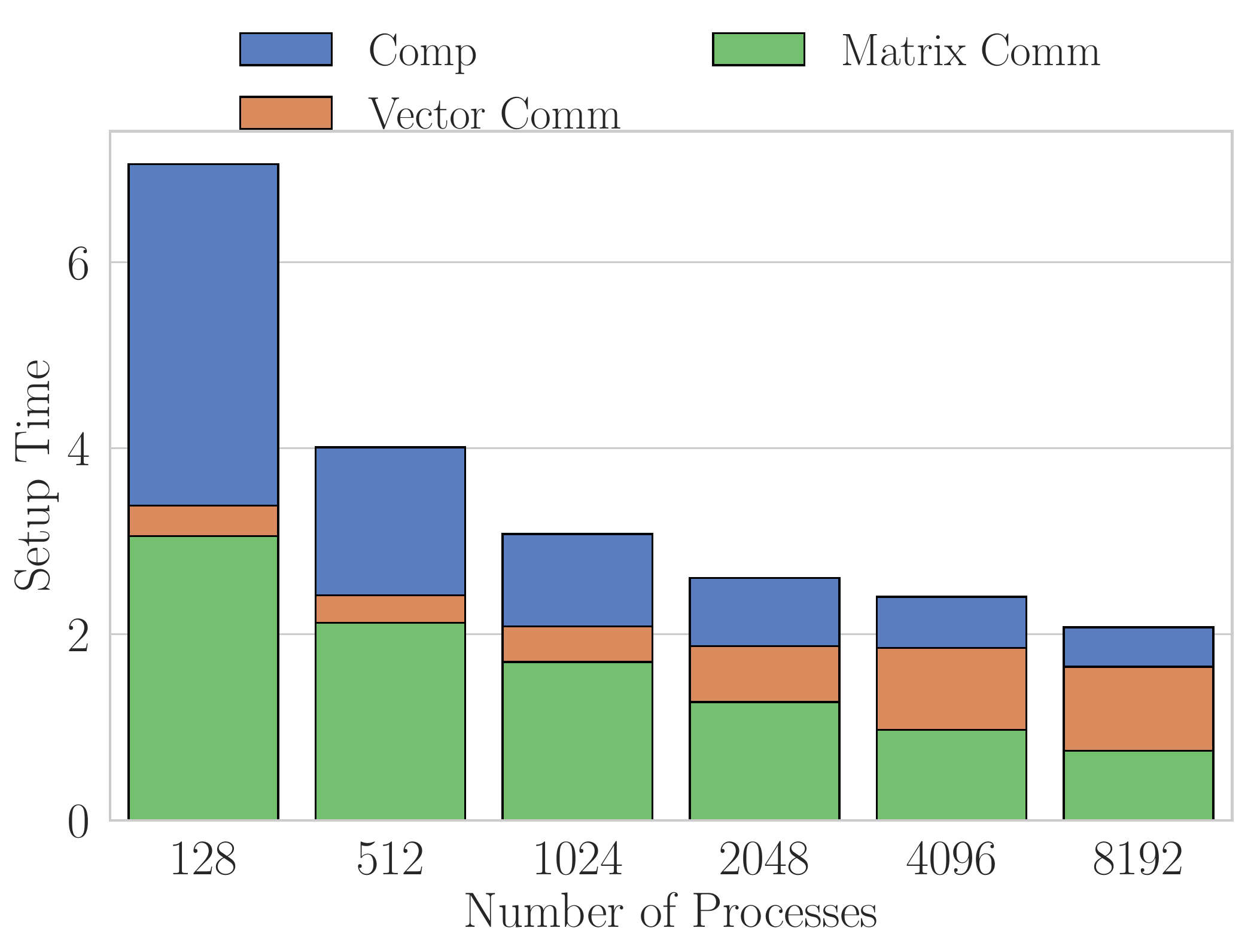}
    \caption{The cost of creating a Ruge-St\"{u}ben hierarchy for
    Example~\ref{exmp:laplace} on various core counts.}\label{figure:setup_comm_scale}
    \Description{Setup Comm Cost, Scaled}
\end{figure}

Point-to-point communication dominates the total cost of the setup phase,
particularly when a large number of processes are active in construction of the
hierarchy.  Figure~\ref{figure:setup_comm_scale} displays the cost of forming a
Ruge-St\"{u}ben hierarchy for Example~\ref{exmp:laplace}, strongly scaled
across a variety of core counts.  As the number of processes is increased, a
larger percentage of time is associated with point-to-point communication.

After the hierarchy is constructed, the solve phase iterates over all levels
until convergence.  This phase, described in Algorithm~\ref{alg:solve} consists
of relaxing error with a smoother such as Jacobi or Gauss-Seidel, calculating
the residual, and restricting this residual to a coarser level where this
process is repeated until error can be solved for directly.  Finally, error
from the coarser level is interpolated up the hierarchy, added to the current
solution, and again smoothed with a relaxation method.

\begin{figure}[ht!]
    \centering
    \begin{subfigure}{0.49\textwidth}
        \centering
        \includegraphics[width=0.75\linewidth]{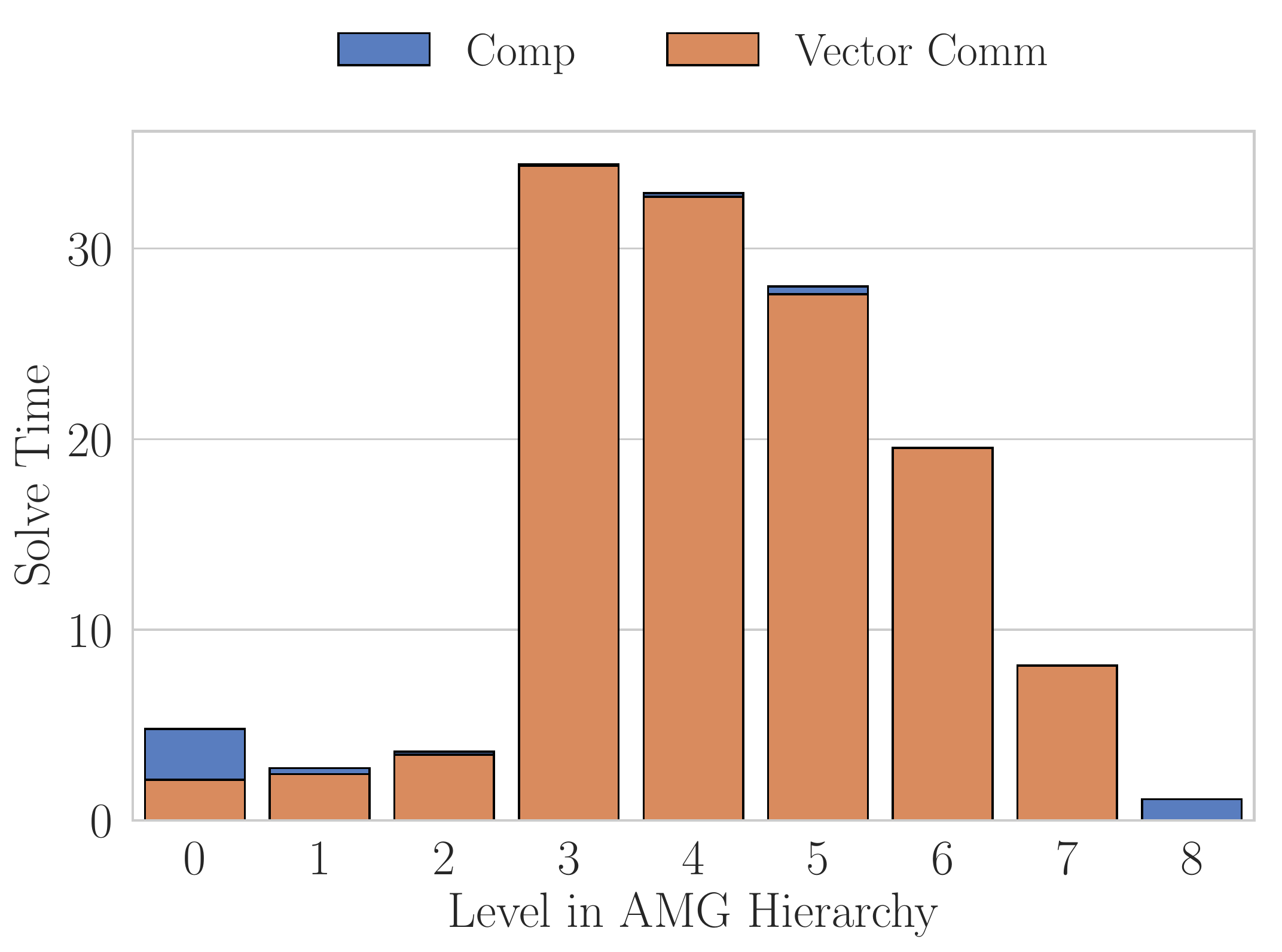}
        \caption{Ruge-Stuben}\label{figure:profile_rss_comm_solve}
    \end{subfigure}
        \begin{subfigure}{0.49\textwidth}
        \centering
        \includegraphics[width=0.75\linewidth]{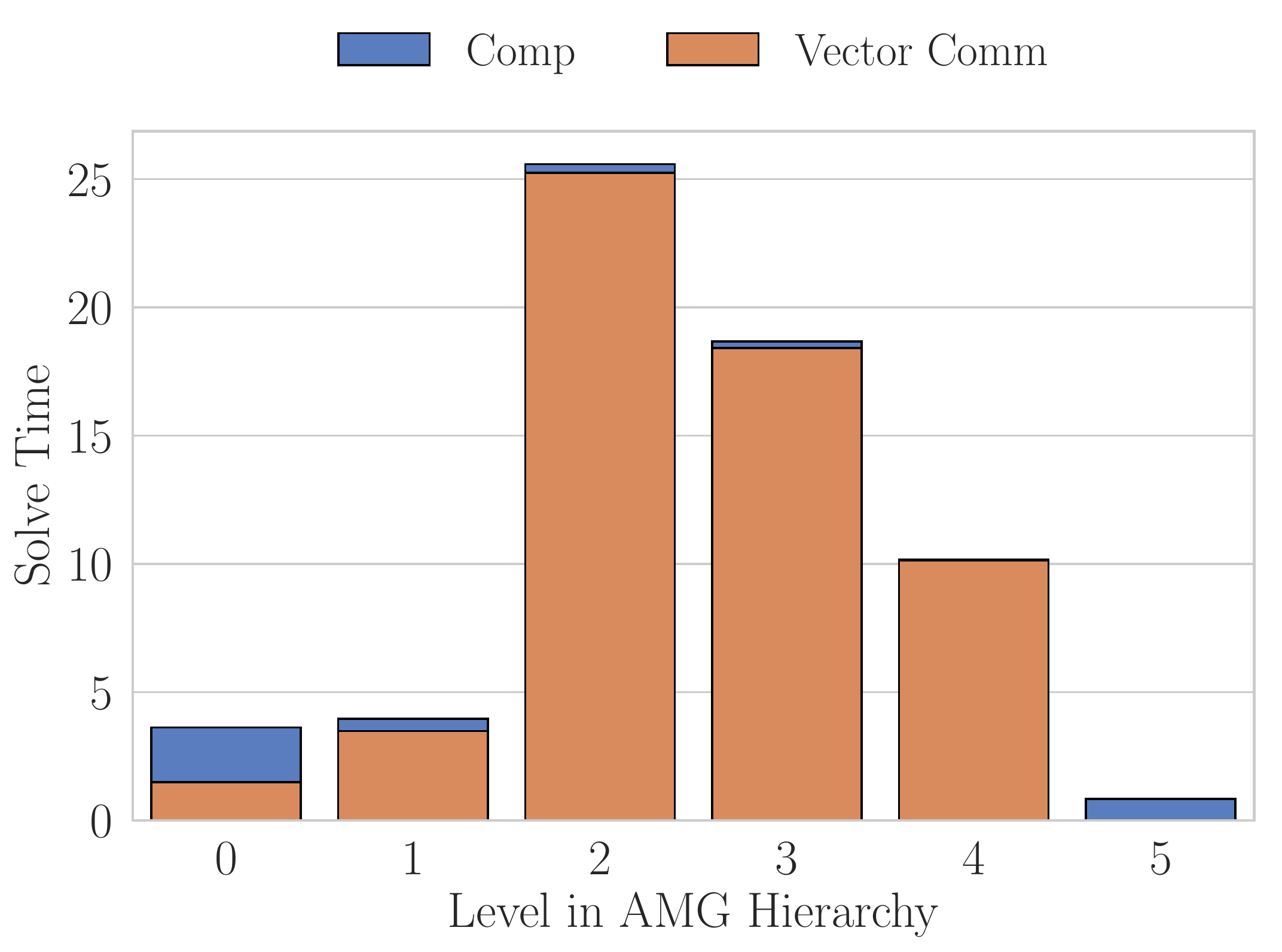}
        \caption{Smoothed Aggregation}\label{figure:profile_sas_comm_solve}
    \end{subfigure}
    \caption{The solve phase cost for various AMG hierarchies for the system from
    Example~\ref{exmp:laplace}.  The total cost is partitioned by level, and
    further split into communication and local computation.}\label{figure:solve_comm}
    \Description{Solve Comm Cost}
\end{figure}

Figure~\ref{figure:solve_comm} displays the cost of iteratively solving the
Ruge-St\"{u}ben and smoothed aggregation hierarchies for
Example!\ref{exmp:laplace}.  The solve phase costs are partitioned into local
computation and MPI vector communciation, associating the large increase in
cost on coarse levels with point-to-point communication.

\begin{figure}[h]
    \centering
    \includegraphics[width=0.375\textwidth]{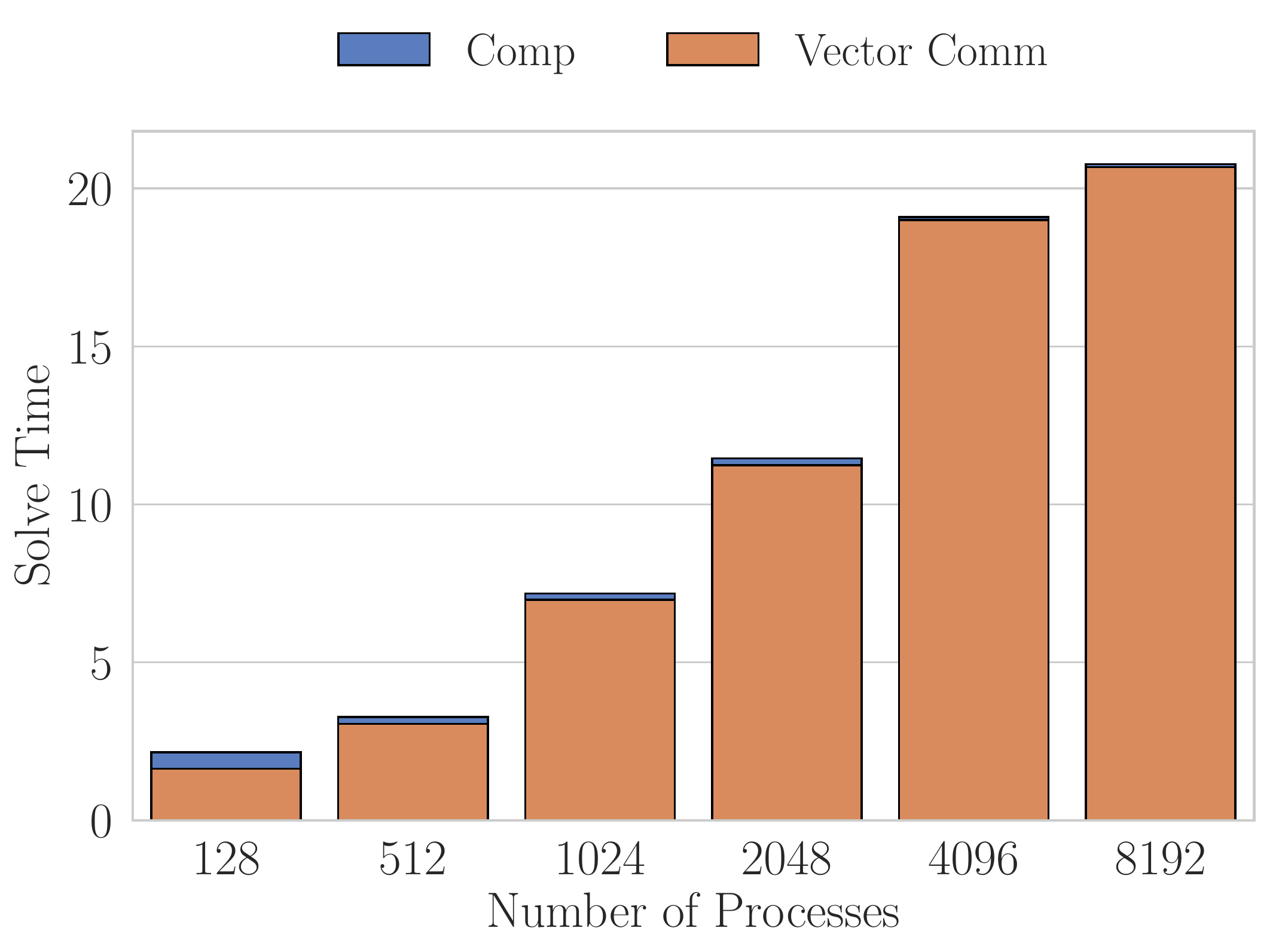}
    \caption{The cost of iteratively solving the Ruge-St\"{u}ben hierarchies for the system from
    Example~\ref{exmp:laplace} on various core counts.}\label{figure:solve_comm_scale}
    \Description{Solve Comm Cost, Scaled}
\end{figure}

MPI communication dominates the cost of the algebraic multigrid solve phase,
particularly with increase in scale.  Figure~\ref{figure:solve_comm_scale}
shows the full cost of the iterative solve phase of AMG at various scales.  As
the number of processes increases, the percentage of cost due to
communication also increases, even as the problem size stays constant.

\subsection{Parallel Matrix Operations}
Parallel vector and sparse matrix communication dominates the cost of algebraic
multigrid, particularly at large scales.  This point-to-point communication is
required for parallel sparse matrix operations in methods of both the setup and
solve phases.

\begin{figure}[ht!]
    \captionsetup[subfigure]{justification=centering}
    \centering
    \includegraphics[height=1.2in]{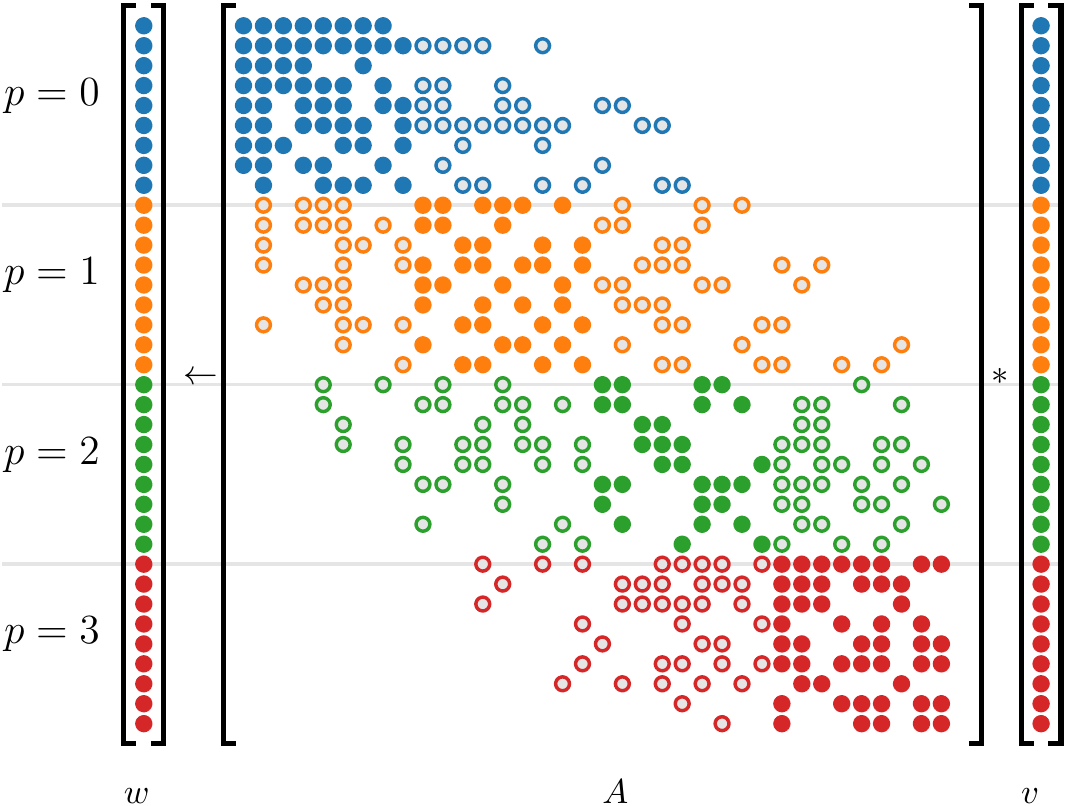}
    \caption{Matrix-vector multiplication across four processes, row-wise.}\label{figure:partition}
    \Description{Parallel SpMV.}
\end{figure}

Assuming a row-wise partition of a linear system, as displayed in
Figure~\ref{figure:partition}, each process holds a contiguous subset of the
rows of the matrix, along with corresponding vector values.
The local rows of the matrix are further split into an on-process block,
associated with local vector values, as well as off-process columns, which
correspond to vector entries stored on other processes.  As a result,
matrix operations such as sparse matrix-vector communication require
communication of vector values corresponding to non-zero off-process columns.
Hence, vector communication consists of each process sending vector values to
any process with corresponding non-zero columns.  This communication pattern is
initialized during construction of the matrix.

\begin{figure}[ht!]
    \captionsetup[subfigure]{justification=centering}
    \centering
    \includegraphics[height=1.2in]{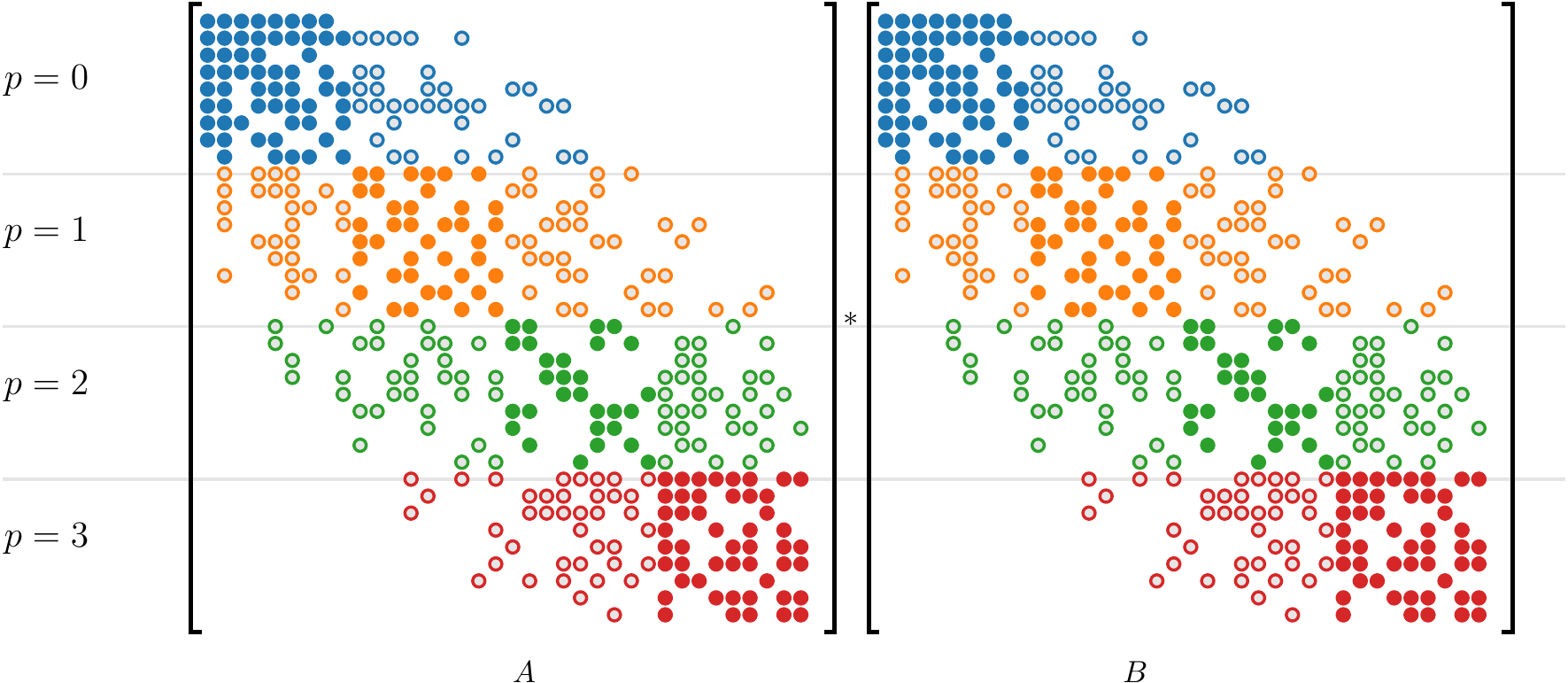}
    \caption{Two matrices partitioned across four processes in a row-wise
    manner.}\label{figure:partition_mat}
    \Description{Parallel SpGEMM}
\end{figure}

Similarly, matrix communication depends on the non-zero off-process
columns.  Figure~\ref{figure:partition_mat} shows two matrices, $A$ and $B$,
partitioned row-wise across four processes, with the local rows again
partitioned into on and off-process columns.
Matrix operations such as sparse matrix-matrix multiplication require
communication of the rows of $B$ that correspond to non-zero off-process columns
of $A$.  In essence, matrix communication retains the same communication pattern
as that of vectors, but requiring entire rows of the matrix rather than
single values.

This \textit{point-to-point} communication~\cite{MPICH} dominates the total time
(both setup and solve) in coarse levels of AMG, as show in
Figure~\ref{figure:amg_scale}.  The cost associated with communication increases
on coarse levels.

\section{Node-Aware Communication}\label{section:node_aware}

The cost associated with standard point-to-point communication throughout AMG
can be reduced through the use of node-awareness, particularly when a large
number of messages are communicated, as is the case on coarse levels of AMG\@.
This concept is introduced in~\cite{Bienz_napspmv} for the SpMV and is extended
here to all components of the AMG setup and solve phases.  In particular, a new
two-step communication process is introduced for the finest levels of the AMG
hierarchy.

\begin{figure}[ht!]
    \captionsetup[subfigure]{justification=centering}
    \centering
    \includegraphics[height=1.2in]{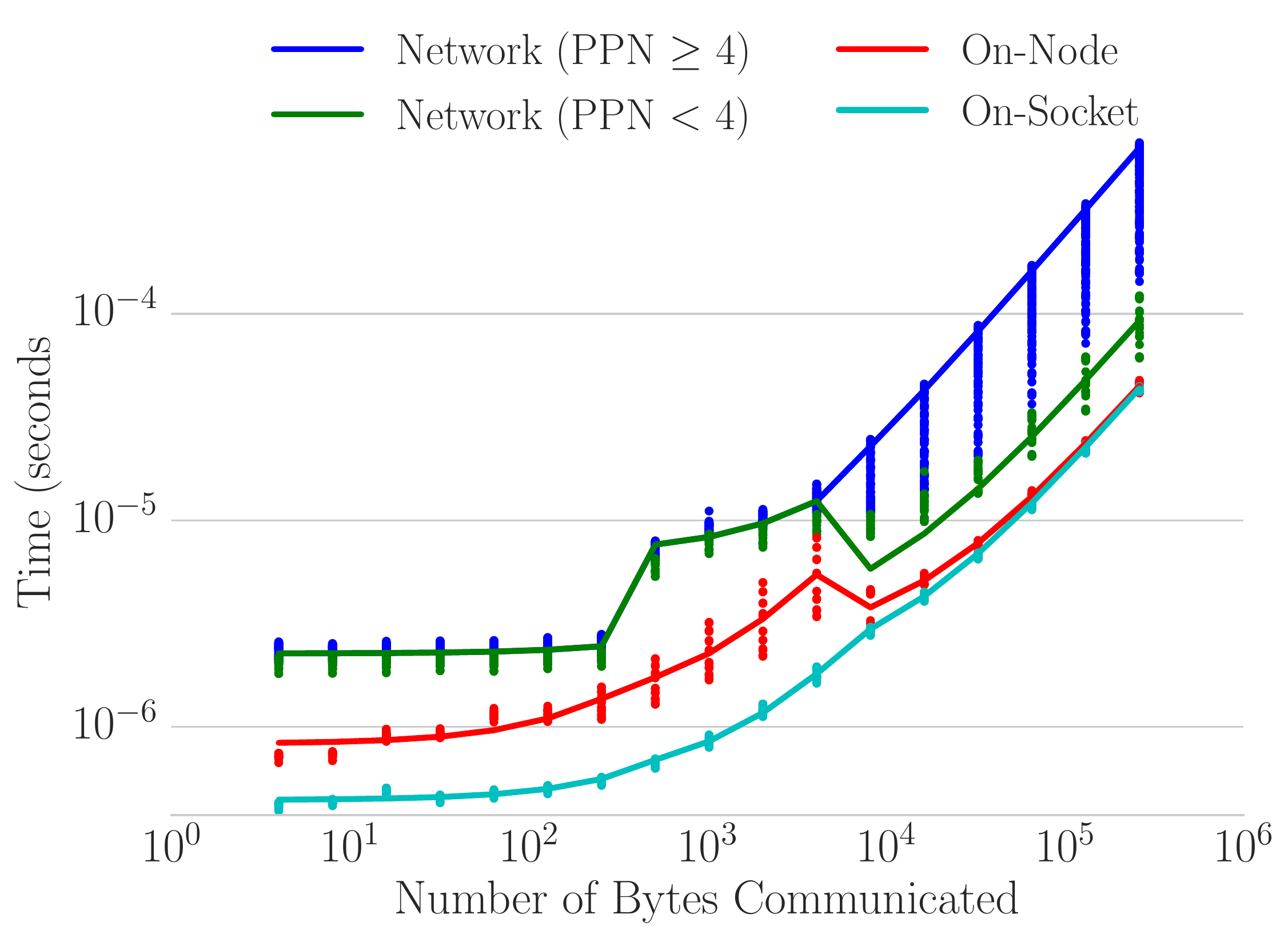}
    \caption{The cost of sending a single message between two
    processes located on the same socket, on different sockets of the same node,
    or on different nodes.  The scattered dots represent timings, while the
    thick lines are corresponding models.}\label{figure:ping_pong}
    \Description{Ping-pong Test.}
\end{figure}

\begin{figure}[ht!]
    \centering
    \includegraphics[height=1.2in]{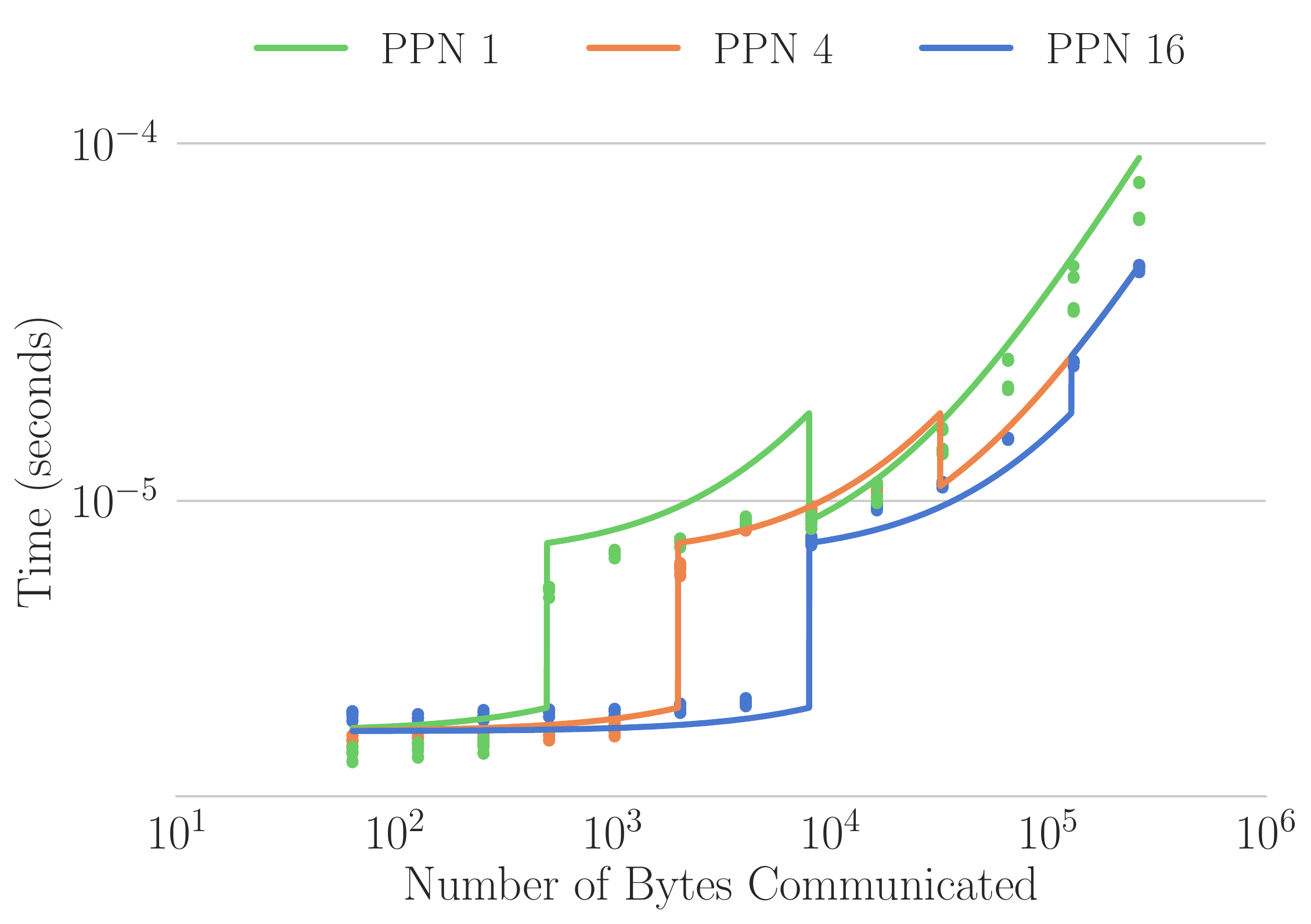}
    \caption{The cost of sending a single message between two processes on
    different nodes.  Various amounts of data are sent between the two nodes,
    with between $1$ and $16$ processes communicating an even portion of this
    data per node.}\label{figure:ping_pong_node}
    \Description{Ping-pong Node Test.}
\end{figure}

The cost of communication depends on many factors, such as number of messages,
size of the messages, and relative locations of the send and receive processes.
For instance, messages between two processes on the same socket are
significantly cheaper than communication between processes located on the
same node but different sockets.  Communication cost is further increased when
the send and receive processes are on different nodes, requiring messages to be
injected into the network.   Figure~\ref{figure:ping_pong} shows the
difference in the cost of sending a single message relative to the location of
participating processes, with measured timings represented as scattered dots
while the corresponding thick lines display the associated model measurements.
The model is calculated with the max-rate model, which adds bandwidth injection
limits to the standard postal model~\cite{MaxRate}, and the ping-pong tests are
measured with Nodecomm\footnote{See \url{https://bitbucket.org/william_gropp/baseenv}}.
Furthermore, the cost of communicating data between nodes is dependent on the
number of active processes, with the cost minimized as data is distributed
across a larger number of processes.  Figure~\ref{figure:ping_pong_node}
displays the cost of sending a single message between two nodes, with various
numbers of active processes.
Additional parameters not included in the max-rate model add to the cost of
communication, such as queue search costs, which result from sending a large
number of messages at once, and network contention, which occurs when many
processes communicate large amounts of data across multiple links of the
network~\cite{BienzEuroMPI}.  The large costs associated with inter-node messages motivate
replacing them with extra intra-node messages when possible.  However, a node's
communication load should remain balanced with all $\texttt{ppn}$ processes sending
approximately $\frac{1}{\texttt{ppn}}^{\textnormal{th}}$ of the data to minimize
bandwidth costs.

\begin{figure}[ht!]
    \captionsetup[subfigure]{justification=centering}
    \centering
    \includegraphics[width=0.375\textwidth]{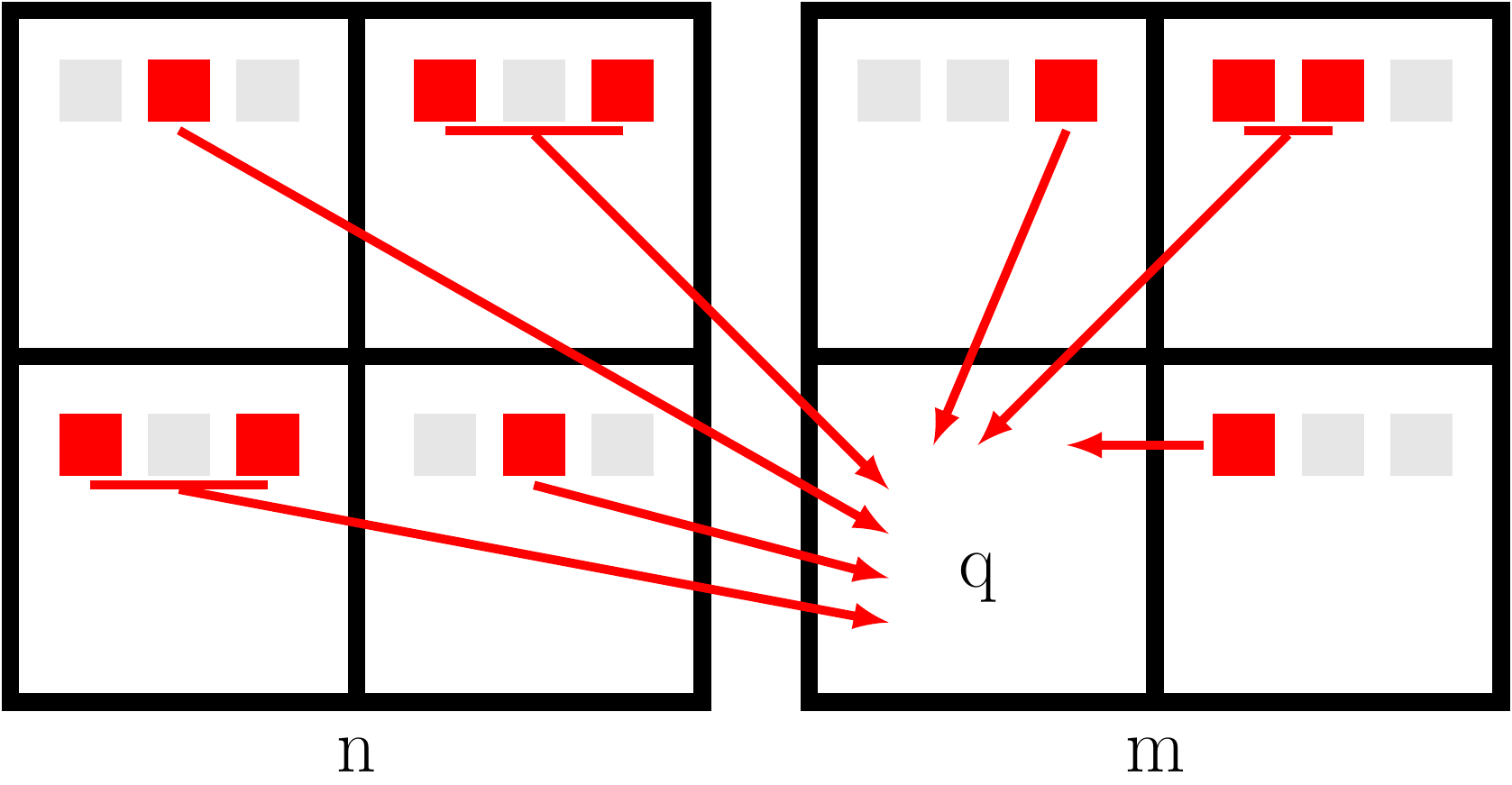}
    \caption{Standard communication between processes on node $n$ and a process
    $q$ on node $m$ yields multiple messages to be sent between the two
    nodes.}\label{figure:standard_comm_a}
    \Description{Standard communication.}
\end{figure}

\begin{figure}[ht!]
    \captionsetup[subfigure]{justification=centering}
    \centering
    \includegraphics[width=0.375\textwidth]{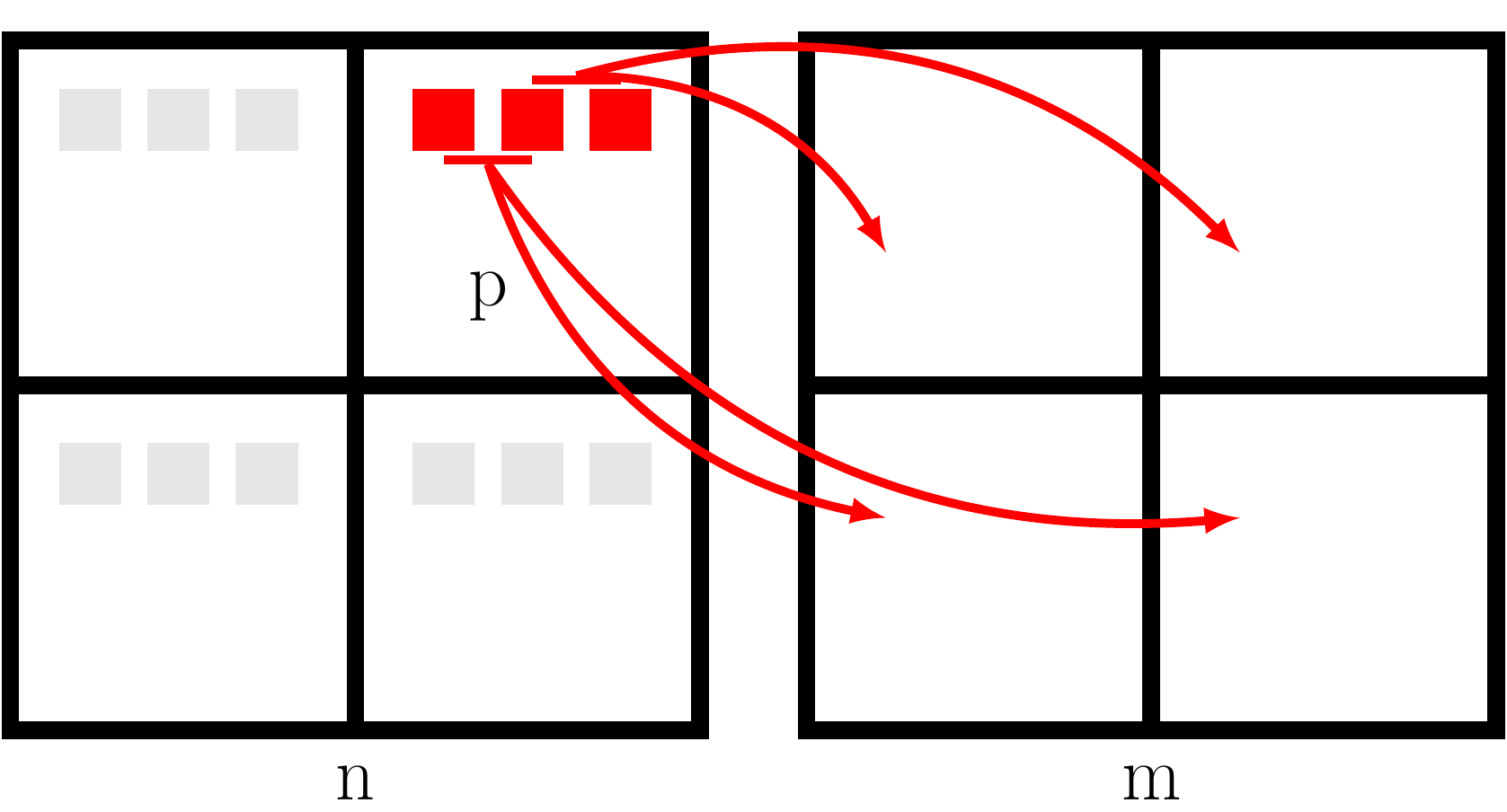}
    \caption{Standard communication from a process $p$ to all processes on node
    $m$ yields multiple messages to be communicated between nodes $n$ and $m$,
    while also sending duplicate data through the network.}\label{figure:standard_comm_b}
    \Description{Standard communication.}
\end{figure}

Standard communication requires sending data directly between
processes, regardless of their locations within the parallel topology.  For
example, Figure~\ref{figure:standard_comm_a} displays standard communication in
which a number of processes on nodes $n$ and $m$ send data directly to a process
$q$.
Furthermore, Figure~\ref{figure:standard_comm_b} shows the standard
process of communicating data from some process $p$ on node $n$ to all processes
on node $m$.
In both cases, multiple messages are communicated between the two nodes.
Furthermore, in the latter example, duplicate data is sent to multiple processes
on node $m$, indicating both the number and size of messages communicated
between nodes $n$ and $m$ is larger than ideal.

\subsection{Three-Step Node-Aware Communication}\label{section:three_step}
\begin{figure}[ht!]
    \captionsetup[subfigure]{justification=centering}
    \centering
    \includegraphics[width=0.375\textwidth]{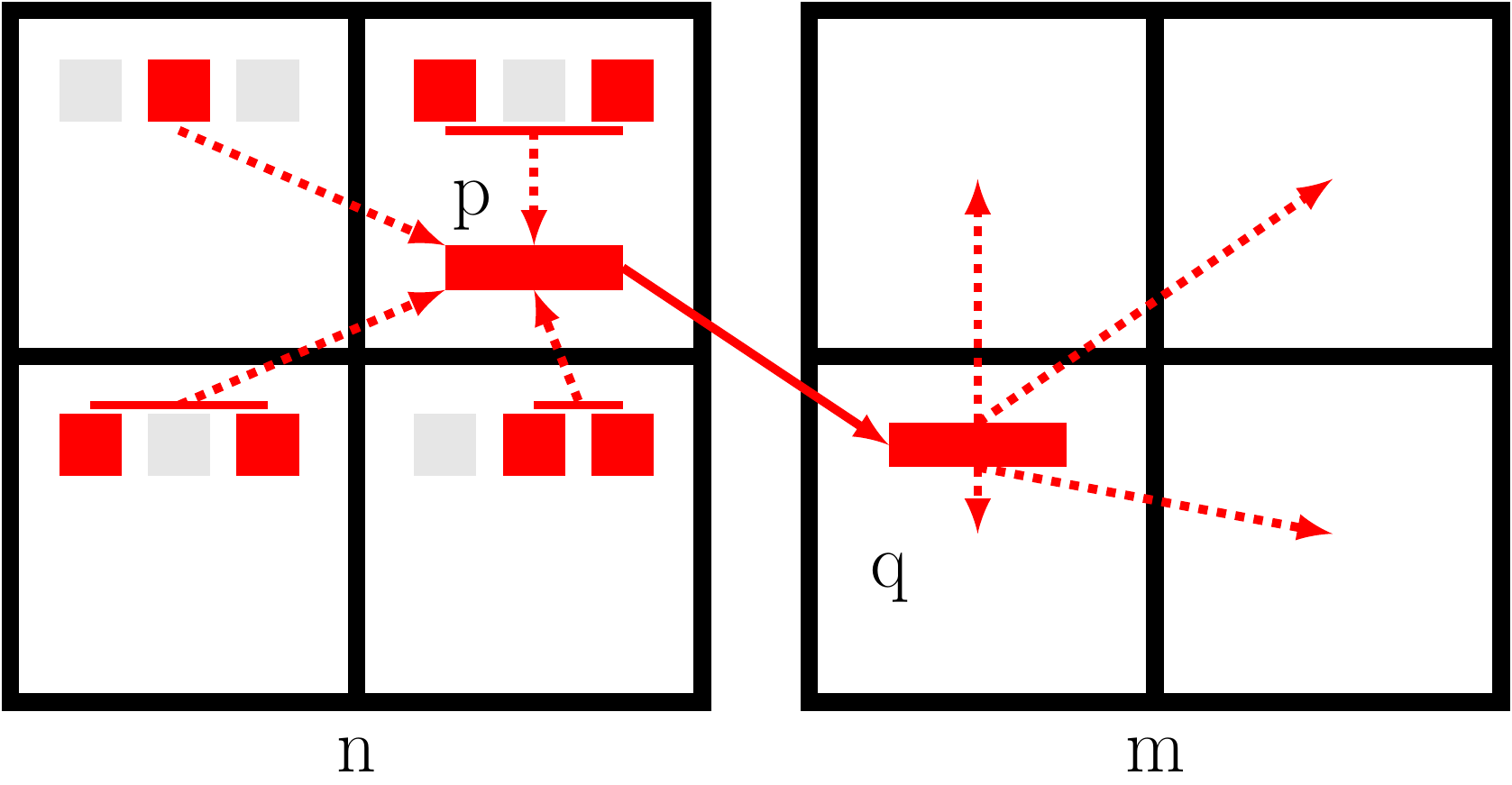}
    \caption{Three-step node-aware communication: (1) gather
      data on a process $p$ local to the node on which data originates, (2)
    send a single message between process $p$ on node $n$ and process
    $q$ on node $m$, and (3) redistribute across processes on node
    $m$.}\label{figure:NAP_three_step}
    \Description{Three-step communication.}
\end{figure}

Three-step node-aware parallel (NAP-3) communication reduces the number and size
of messages injected into the network while increasing the amount of less-costly
on-node communication~\cite{Bienz_napspmv}.  NAP-3 communication gathers all
data to be sent to node $m$ on some process local to the node $n$ on which it
originates.  This data is then sent as a single message through the network,
before being distributed to the necessary processes on node $m$.
Figure~\ref{figure:NAP_three_step} displays the steps of sending data from node
$n$ to $m$, sending first to process $p$ on node $n$.
A single message is then sent from process $p$ on node $n$ to process $q$ on
node $m$.  Finally, process $q$ distributes the received data to processes on
node $m$ that need it.  In addition, all on-node messages, or those for which
the process of origin lies on the same node as the destination process, are
communicated with the standard approach.  It is important to note that it is
likely that there are other nodes in the network to which processes on node $n$
must also send.  This data is gathered on some process $s$ on $n$ that is not
$p$.  As a result, several processes per node are communicating.

NAP-3 communication greatly reduces both the number of messages as well as the
number of bytes injected into the network by any node.  When a node $n$ is
communicating similar amounts of data to many other nodes, the per-process
message size is also greatly reduced.  However, in the case that node $n$ is
communicating the majority of data to a single node $m$, a large imbalance can
occur in communication requirements of the processes local to node $n$, reducing
bandwidth and increasing message cost.

\subsection{Two-Step Node-Aware Communication}\label{section:two_step}
\begin{figure}[ht!]
    \captionsetup[subfigure]{justification=centering}
    \centering
    \includegraphics[width=0.375\textwidth]{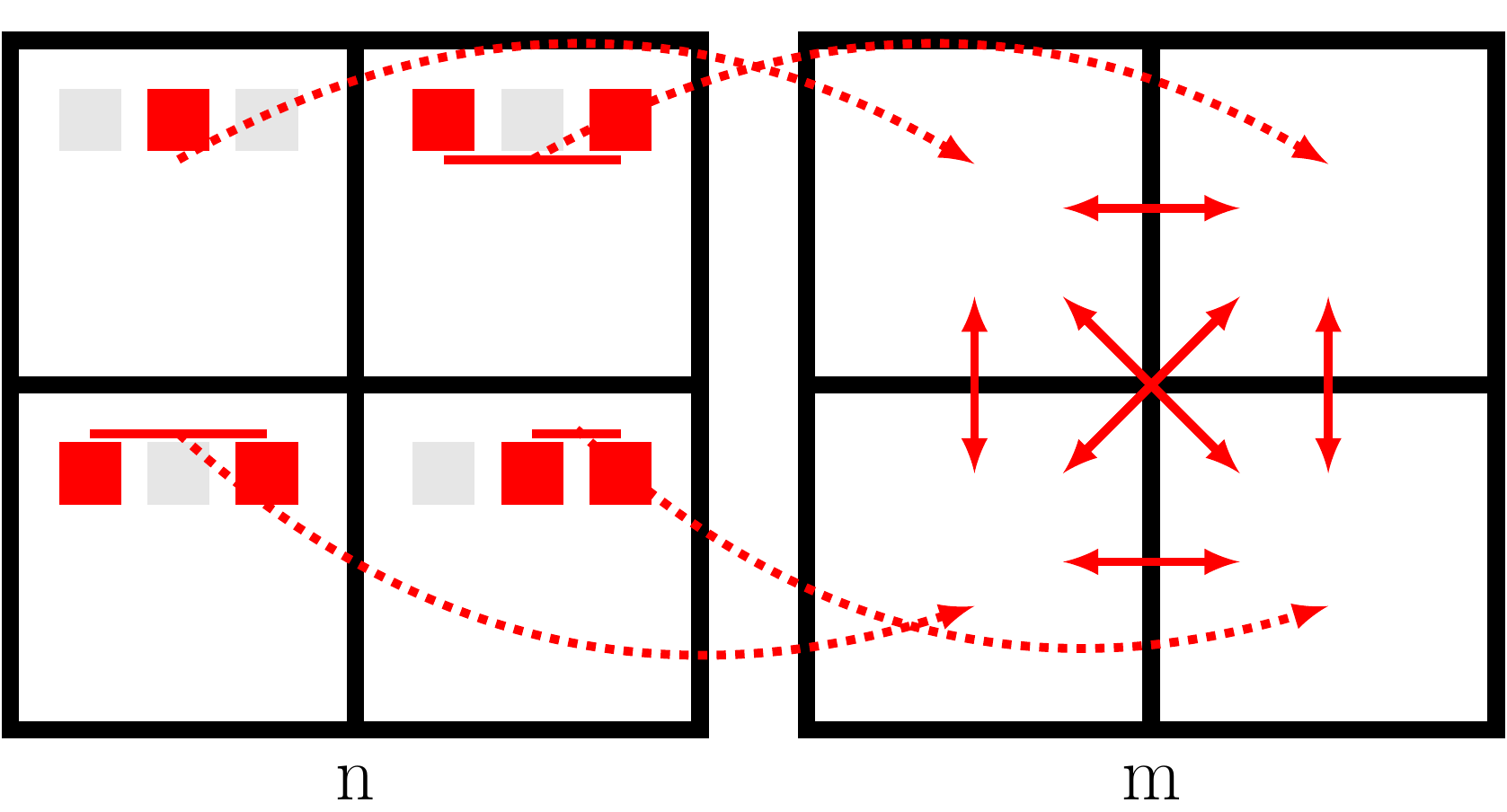}
    \caption{An alternative, two-step node-aware communication: (1)
      gather all data to be sent to node $m$ on process and (2) send
    directly to the corresponding process on node $m$.}\label{figure:NAP_two_step}
    \Description{Alternative communication.}
\end{figure}

Alternatively, this paper introduces a method to allow \textit{all} processes to
remain active in inter-node communication.  This two-step approach to node-aware
communication (NAP-2), displayed in Figure~\ref{figure:NAP_two_step}, consists of
gathering all data on-process to be sent to a node $m$, and sending this
directly to the corresponding process.
This is followed by redistribution of values on the receiving node.  This
alternate node-aware method reduces the number and size of data by eliminating
the duplication displayed in Figure~\ref{figure:standard_comm_b}, but the
multiple messages communicated between nodes in
Figure~\ref{figure:standard_comm_a} remains.  Therefore, NAP-2 communication can
greatly reduce the number and size of inter-node messages over standard
communication while process loads remain equally balanced to standard
communication.  However, as up to $\texttt{ppn}$ messages remain between each set of
nodes, NAP-2 does not reduce the message count to the extent of NAP-3, and
as a result is less beneficial when there is communication between a large
number of nodes.

\subsection{Performance Models for Communication Strategies}\label{section:perf}
\begin{figure*}[ht!]
    \captionsetup[subfigure]{justification=centering}
    \centering
    \begin{subfigure}{0.49\textwidth}
        \centering
        \includegraphics[width=0.75\linewidth]{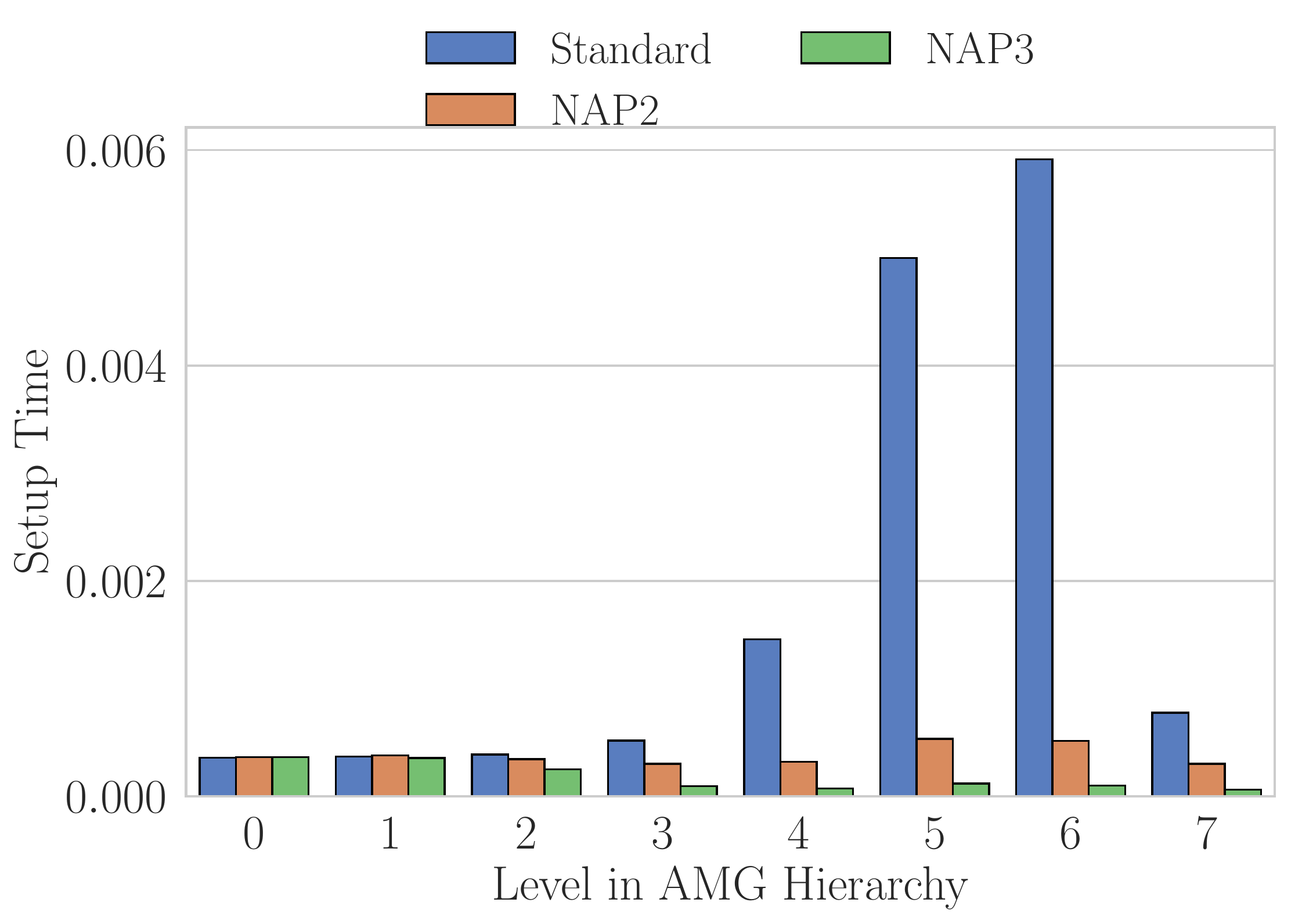}
        \caption{$A \cdot x$}\label{figure:Ax_compare}
    \end{subfigure}
        \begin{subfigure}{0.49\textwidth}
        \centering
        \includegraphics[width=0.75\linewidth]{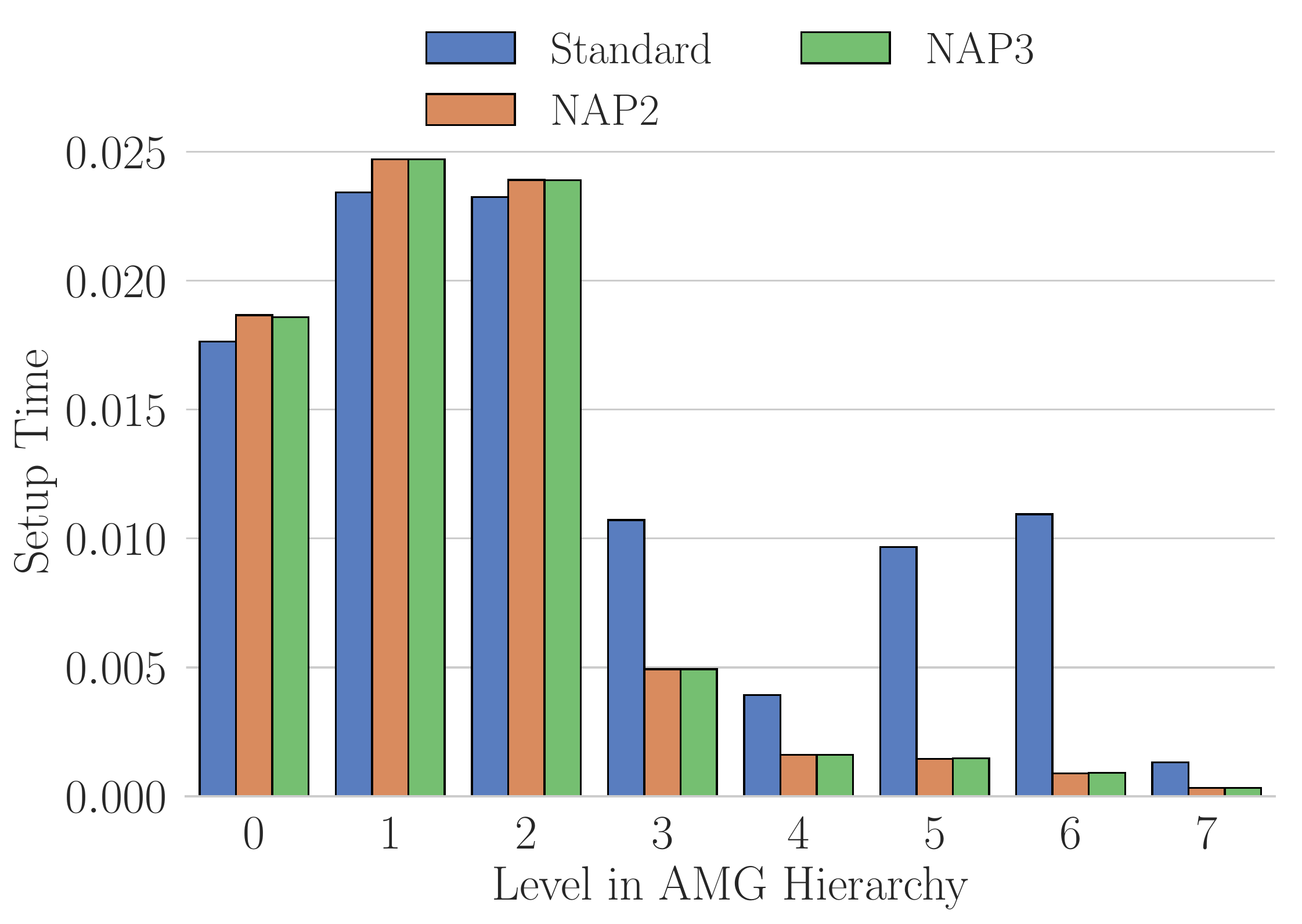}
        \caption{$AP \leftarrow A \cdot P$}\label{figure:AP_compare}
    \end{subfigure}
                            \caption{Cost of standard, two-step node-aware, and three-step
    node aware communication for SpGEMM and SpMV operations from
    Example~\ref{exmp:laplace}}\label{figure:nap_compare}
    \Description{Communication comparison.}
\end{figure*}

The optimal communication strategy varies with problem type, problem scale,
level in AMG hierarchy, and operation.  Figure~\ref{figure:nap_compare} displays
the cost of performing various dominate matrix operations on each level of the
Ruge-St\"{u}ben AMG hierarchy for the Laplace system from
Example~\ref{exmp:laplace}.
\begin{figure*}[ht!]
    \captionsetup[subfigure]{justification=centering}
    \centering
    \begin{subfigure}{0.49\textwidth}
        \centering
        \includegraphics[width=0.75\linewidth]{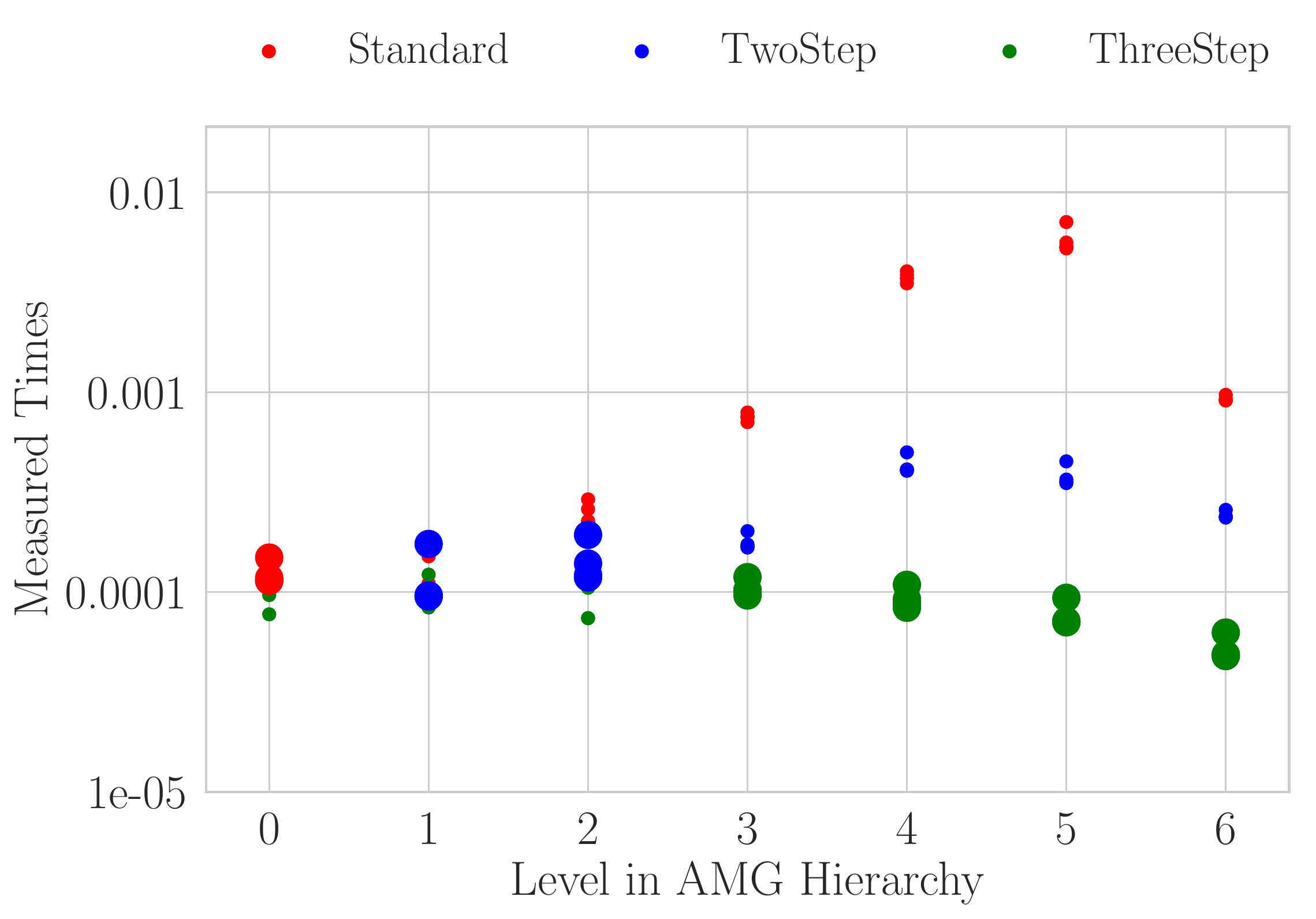}
        \caption{$A \cdot x$}\label{figure:model_Ax}
    \end{subfigure}
        \begin{subfigure}{0.49\textwidth}
        \centering
        \includegraphics[width=0.75\linewidth]{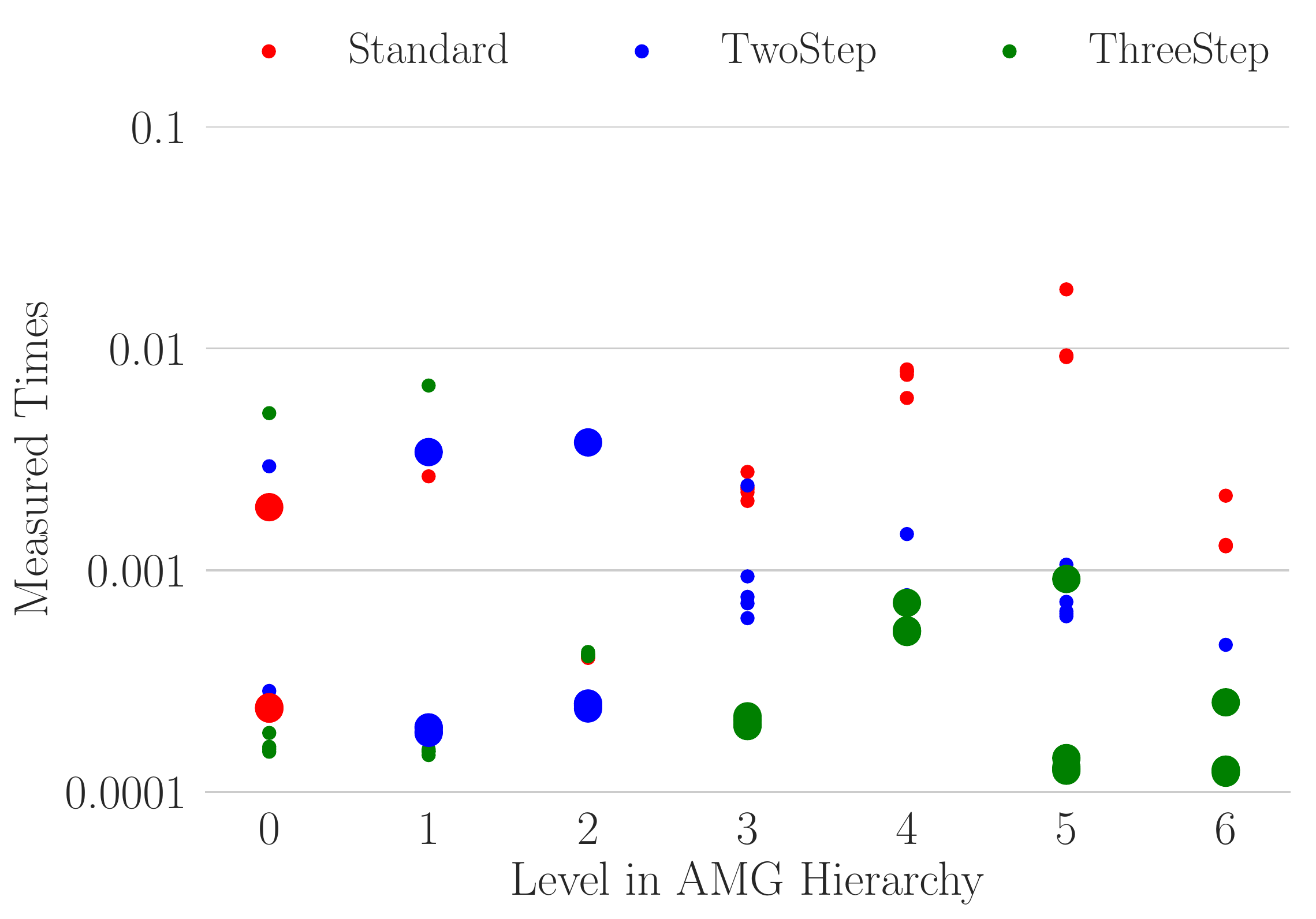}
        \caption{$AP \leftarrow A \cdot P$}\label{figure:model_AP}
    \end{subfigure}
                            \caption{The cost of various AMG operations throughout the Ruge-St\"{u}ben
    hierarchy for Example~\ref{exmp:laplace}.  Each plot contains the operation
    cost associated with standard, NAP2, and NAP3 communication.  The larger
    dots represent the method chosen as minimal by the performance
    models.}\label{figure:model_compare}
    \Description{Models.}
\end{figure*}
While node-aware communication often yields large speedups on coarse levels,
additional work and load imbalance can significantly slowdown fine-level
communication.  Therefore, a performance model should be used to determine which
communication strategy is ideal for the various operations throughout AMG.

The max-rate model~\cite{MaxRate} describes the cost of sending messages from a
symmetric multiprocessing (SMP) node as
\begin{equation}
    T = \alpha n + \frac{\texttt{ppn} s}{\min\left( R_{N}, \texttt{ppn} R_{b}\right)},
    \label{eqn:max_rate}
\end{equation}
where $\alpha$ is the latency, $R_{b}$ is rate at which a process can transport
data, and $R_{N}$ is the rate at which a network interface device (NID) can
inject data into the network.  Furthermore, $n$ is the maximum number of
messages sent and $s$ is the maximum number of bytes.  This model
assumes that all processes on a node communicate equal amounts of data.
However, as NAP-3 can result in imbalanced communication loads, the max-rate
model is altered to be
\begin{equation}
    T = \alpha n + \max\left( \frac{s_{\texttt{node}}}{R_{N}},
    \frac{s_{\texttt{proc}}}{R_{b}}
    \right),\label{eqn:max_rate_adjusted}
\end{equation}
where $s_{\texttt{proc}}$ is the maximum number of bytes communicated by any
process and $s_{\texttt{node}}$ is the maximum number of bytes injected by any NID.
In the case of perfect load balance, $s_{\texttt{node}}$ is equal to
$\texttt{ppn} \cdot s_{\texttt{proc}}$, and Equation~\ref{eqn:max_rate_adjusted}
reduces to the original max-rate model.  Furthermore, when modeling the cost of
intra-node communication, the max-rate model reduces to the standard postal
model
\begin{equation}
    T = \alpha_{\ell} n + \frac{s}{R_{b_{\ell}}},\label{eqn:postal}
\end{equation}
as data is not injected into the network.  In this model, $\alpha_{\ell}$ is
the latency required to send a message to another process on-node, and
$R_{b_{\ell}}$ is the rate at which data is transported between two on-node
processes.  In all models, the latency and bandwidth terms are measured and
applied separately to short, eager, and rendezvous protocols.

The cost of each communication strategy can be approximated as the sum of the
cost of inter-node messages, modeled by Equation~\ref{eqn:max_rate_adjusted},
and intra-node message cost as determined by Equation~\ref{eqn:postal}.
Furthermore, as fully intra-node communication is performed equivalently in
all strategies, only inter-node communication and the node-aware approaches'
additional intra-node communication requirements are modeled.
\begin{figure*}[ht!]
    \captionsetup[subfigure]{justification=centering}
    \centering
    \begin{subfigure}{0.49\textwidth}
        \centering
        \includegraphics[width=0.75\linewidth]{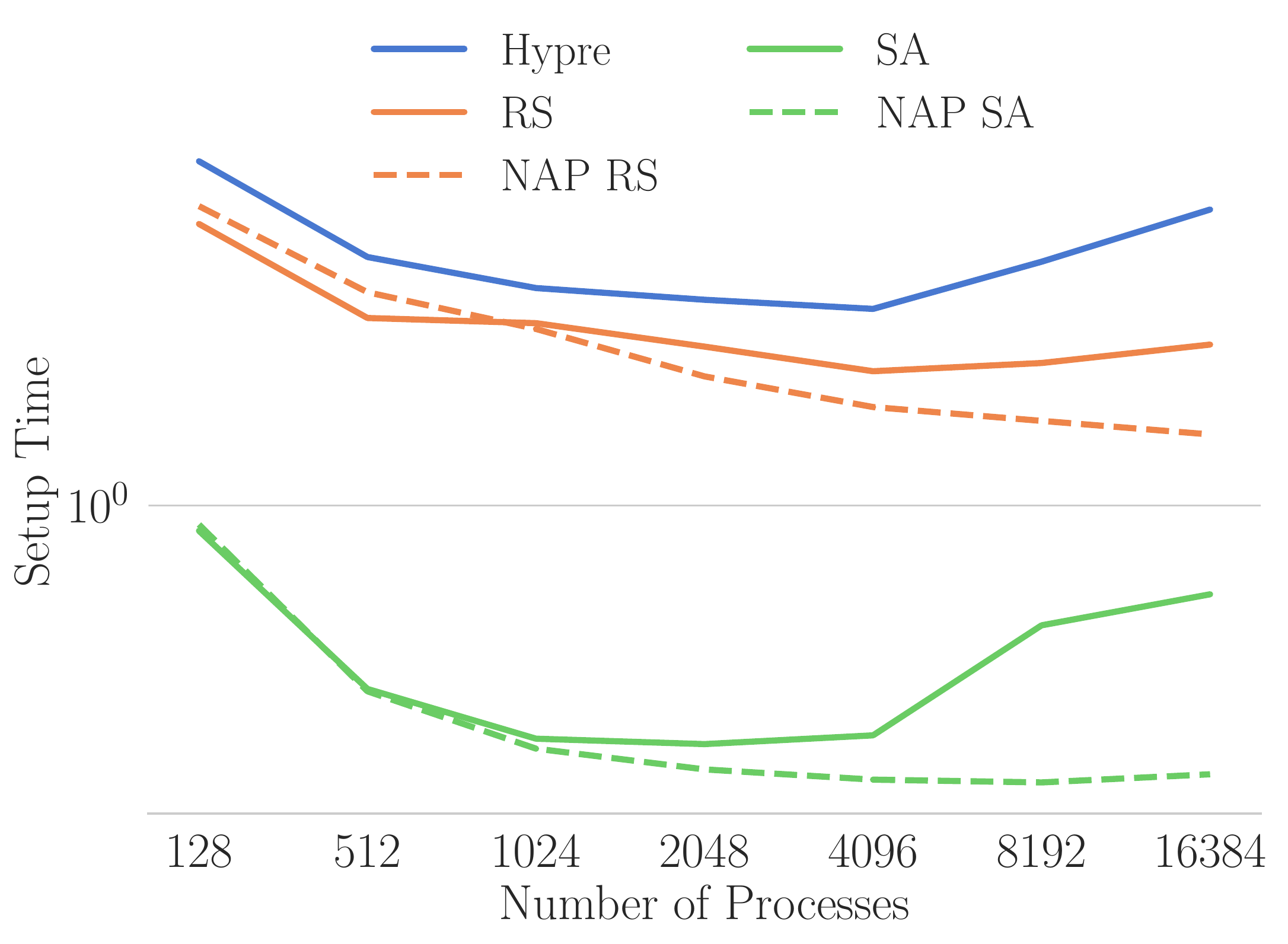}
        \caption{\texttt{Setup Cost}}
    \end{subfigure}    \begin{subfigure}{0.49\textwidth}
        \centering
        \includegraphics[width=.75\linewidth]{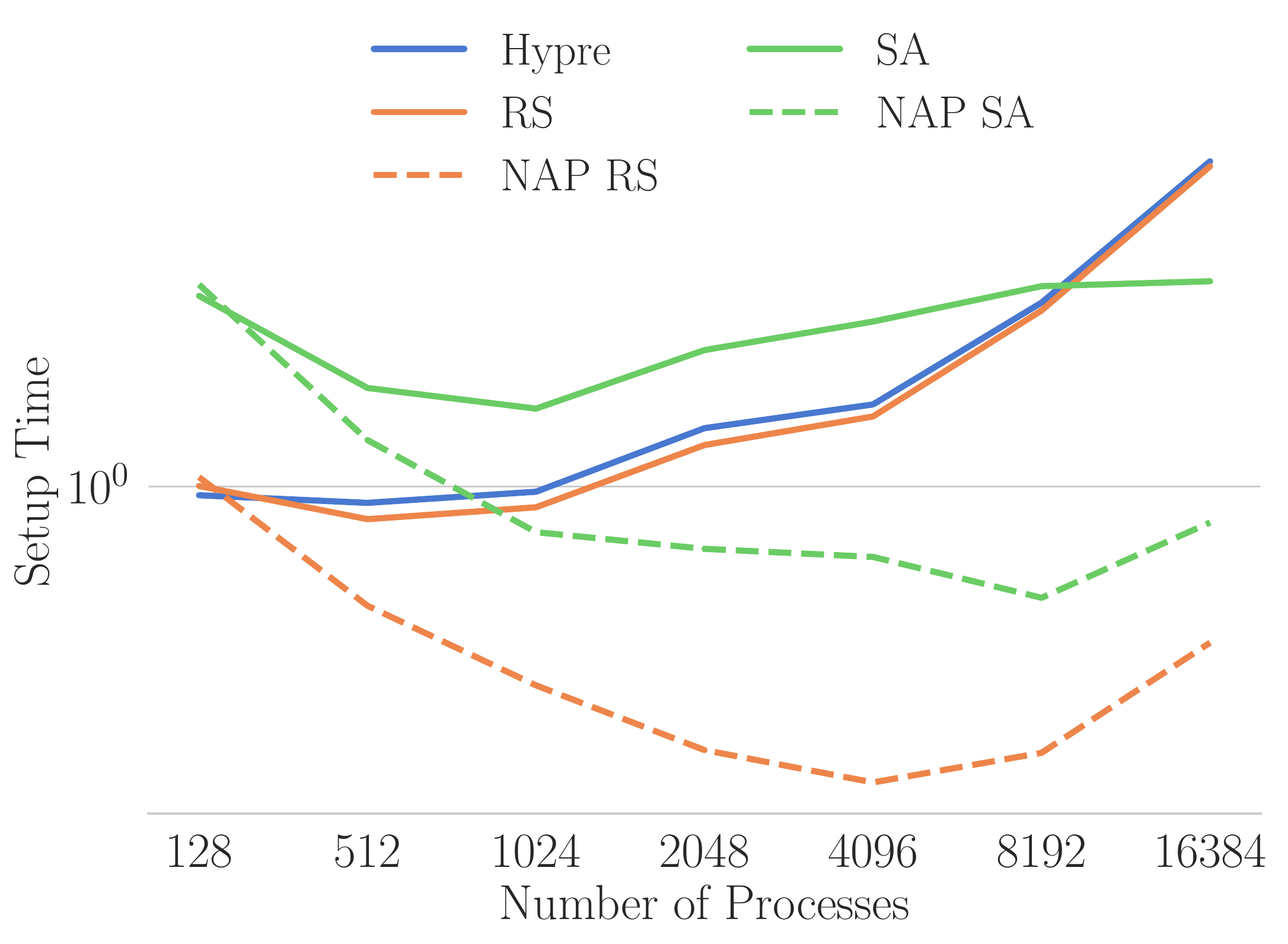}
        \caption{\texttt{Solve Cost}}
    \end{subfigure}
    \caption{Setup and solve times for both the Ruge-St\"{u}ben and Smoothed
    Aggregation hierarchies for the {\bf MFEM Grad-Div}
    system.}\label{figure:mfem_laplace_setup_solve}
    \Description{AMG setup and solve times.}
\end{figure*}

Standard communication is modeled as
\begin{equation}
    T = \alpha n_{\texttt{proc}} +
    \max\left( \frac{s_{\texttt{node}}}{R_{N}},
    \frac{s_{\texttt{proc}}}{R_{b}}
    \right),\label{eqn:standard_model}
\end{equation}
where $n_{\texttt{proc}}$ is the maximum number of processes with which any
process communicates.

Similarly NAP-2 communication is modeled by
\begin{equation}
    T = \alpha n_{\texttt{proc2node}}
    + \max\left( \frac{s_{\texttt{node}}}{R_{N}},
      \frac{s_{\texttt{proc}}}{R_{b}}\right)
    + \alpha_{\ell} \texttt{ppn} -1
    + \frac{s_{\texttt{proc}}}{R_{b_{\ell}}}, \label{eqn:NAP2_model}
\end{equation}
where $n_{\texttt{proc2node}}$ is the maximum number of nodes with which any
process communicates.  Additional intra-node communication is modeled with an
upper bound of $\texttt{ppn} - 1$ messages, transferring a total of $s_{\texttt{proc}}$
bytes.  In this worst case, a process sends all received bytes to the
$\texttt{ppn} - 1$ other
processes on the node.

Finally, NAP-3 communication is modeled as
\begin{equation}
    T = \alpha \frac{n_{\texttt{node2node}}}{\texttt{ppn}}
    + \max\left( \frac{s_{\texttt{node}}}{R_{N}},
      \frac{s_{\texttt{node2node}}}{R_{b}}\right)
    + 2 \cdot \left(\alpha_{\ell} \texttt{ppn} -1
    + \frac{s_{\texttt{node2node}}}{R_{b_{\ell}}} \right), \label{eqn:NAP3_model}
\end{equation}
where $n_{\texttt{node2node}}$ and $s_{\texttt{node2node}}$ are the number and
size of messages, respectively, communicated between any two nodes.

These models do not take into account reductions in the size of node-aware
messages that result from removing duplicate data.  Furthermore, the additional
intra-node communication in the NAP-2 and NAP-3 models is a rough estimate of an
upper bound.  However, the models are accurate enough to distinguish between the
various strategies.

Figure~\ref{figure:model_compare} shows the cost of communication in the SpGEMM
$A \cdot P$ and the SpMV $A \cdot x$ on each level of the AMG hierarchy from
Example~\ref{exmp:laplace} for each of the communication strategies.  Each
operation is tested multiple times, and all timings are plotted.  The
enlarged circles represent the communication strategy with minimum modeled
measurement.  While the measurements vary among the runs, the model accurately
chooses a communication strategy that performs at least as well as standard
communication.

\section{Results}\label{section:results}
\begin{figure*}[ht!]
    \captionsetup[subfigure]{justification=centering}
    \centering
    \begin{subfigure}{0.49\textwidth}
        \centering
        \includegraphics[width=0.75\linewidth]{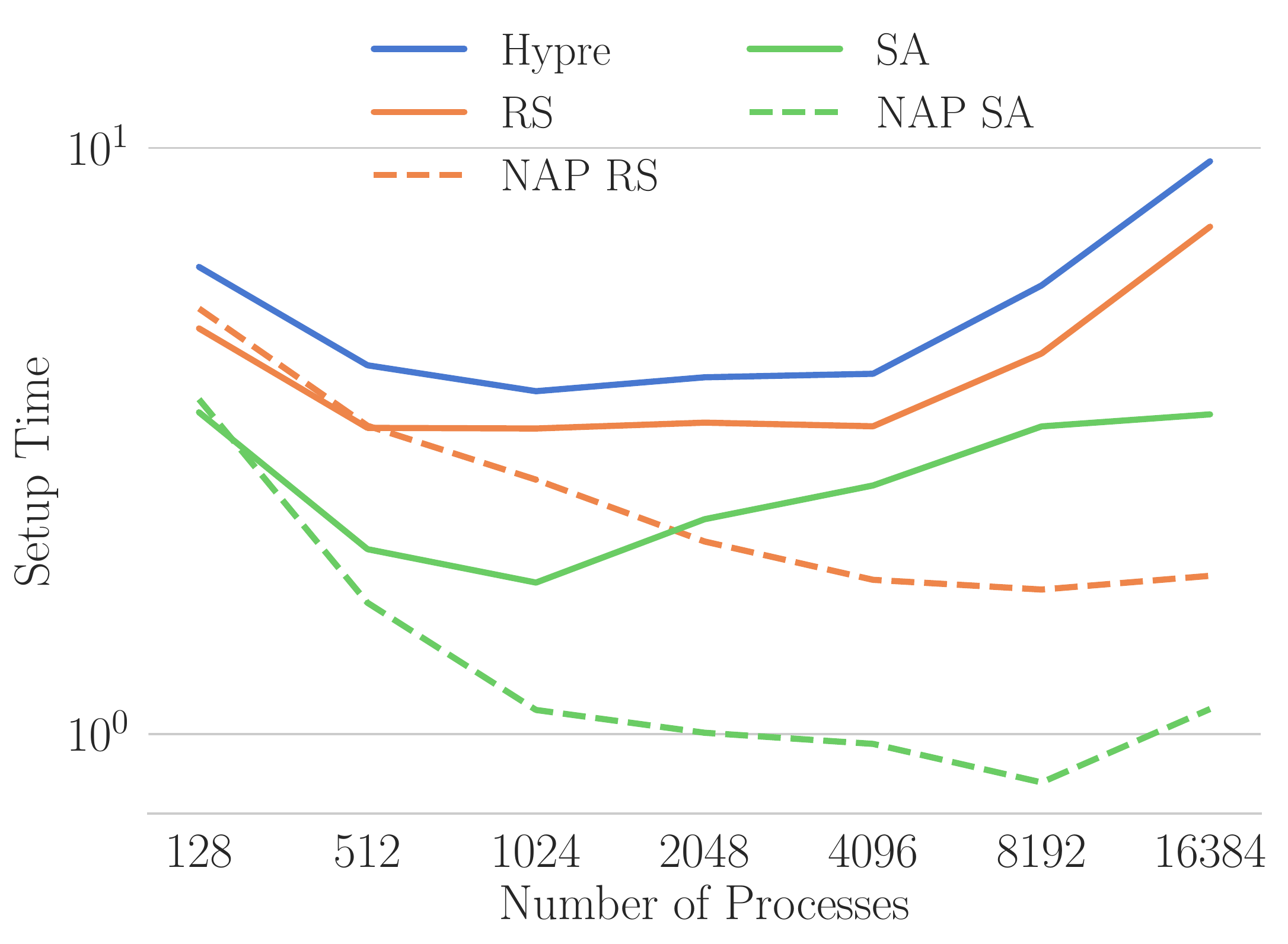}
        \caption{\texttt{Total Cost}}
    \end{subfigure}    \begin{subfigure}{0.49\textwidth}
        \centering
        \includegraphics[width=.75\linewidth]{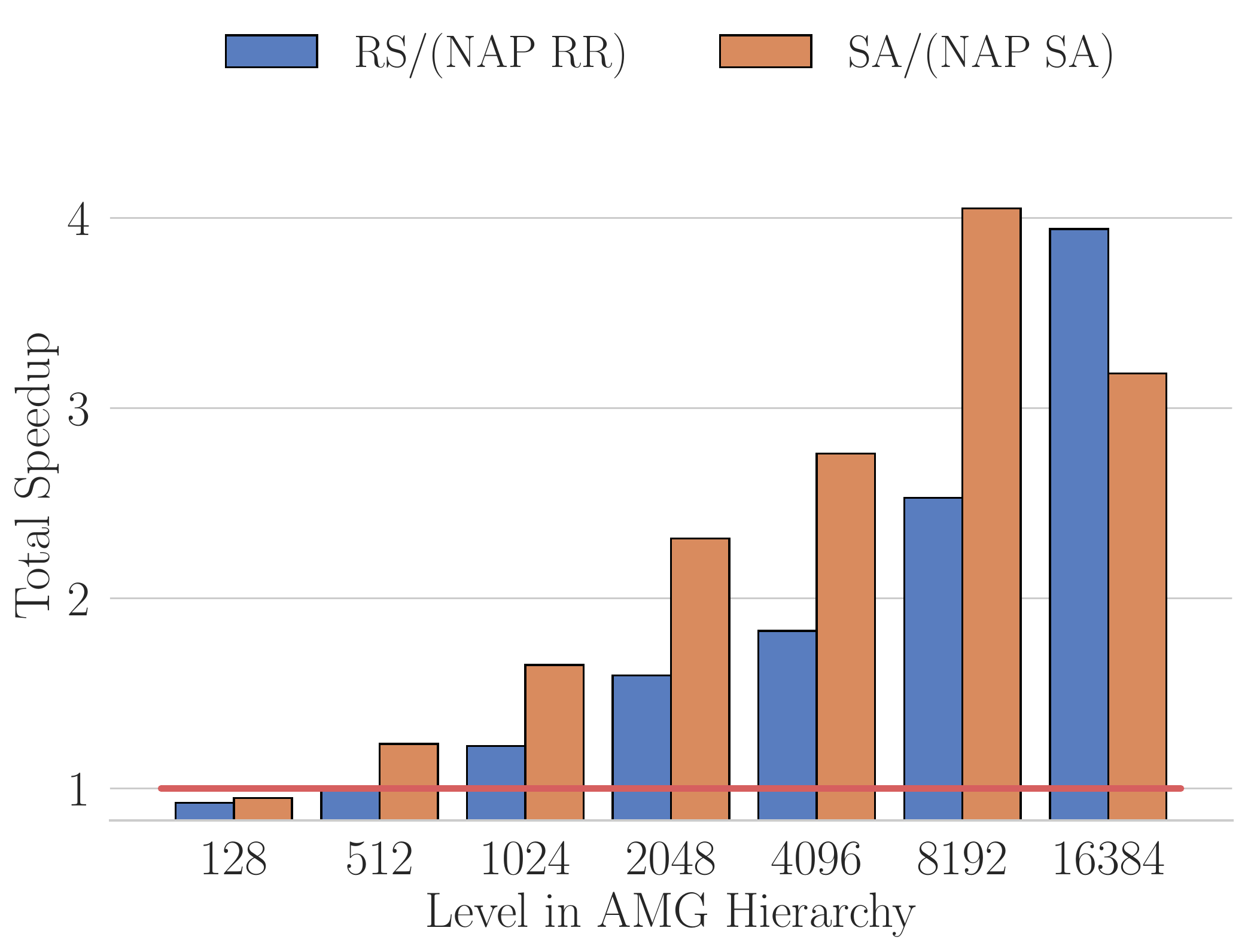}
        \caption{\texttt{Node-Aware Speedup}}
    \end{subfigure}
    \caption{Total AMG times for both the Ruge-St\"{u}ben and Smoothed
    Aggregation hierarchies for {\bf MFEM Grad-Div} on Blue
    Waters.}\label{figure:mfem_grad_div_total_time}
    \Description{Total GradDiv AMG Times on Blue Waters.}
\end{figure*}
\begin{figure*}[ht!]
    \captionsetup[subfigure]{justification=centering}
    \centering
    \begin{subfigure}{0.49\textwidth}
        \centering
        \includegraphics[width=0.75\linewidth]{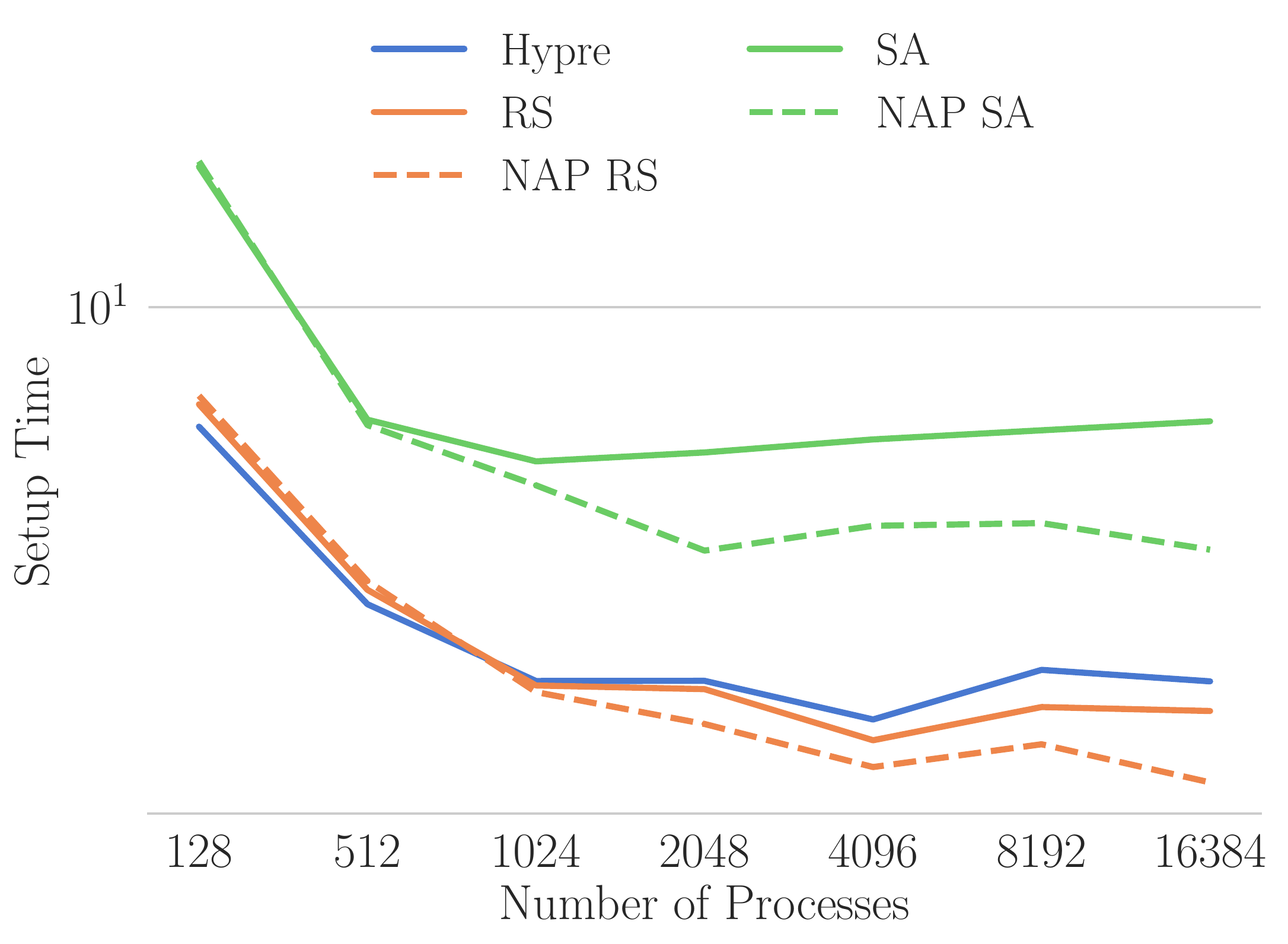}
        \caption{\texttt{Total Cost}}
    \end{subfigure}    \begin{subfigure}{0.49\textwidth}
        \centering
        \includegraphics[width=.75\linewidth]{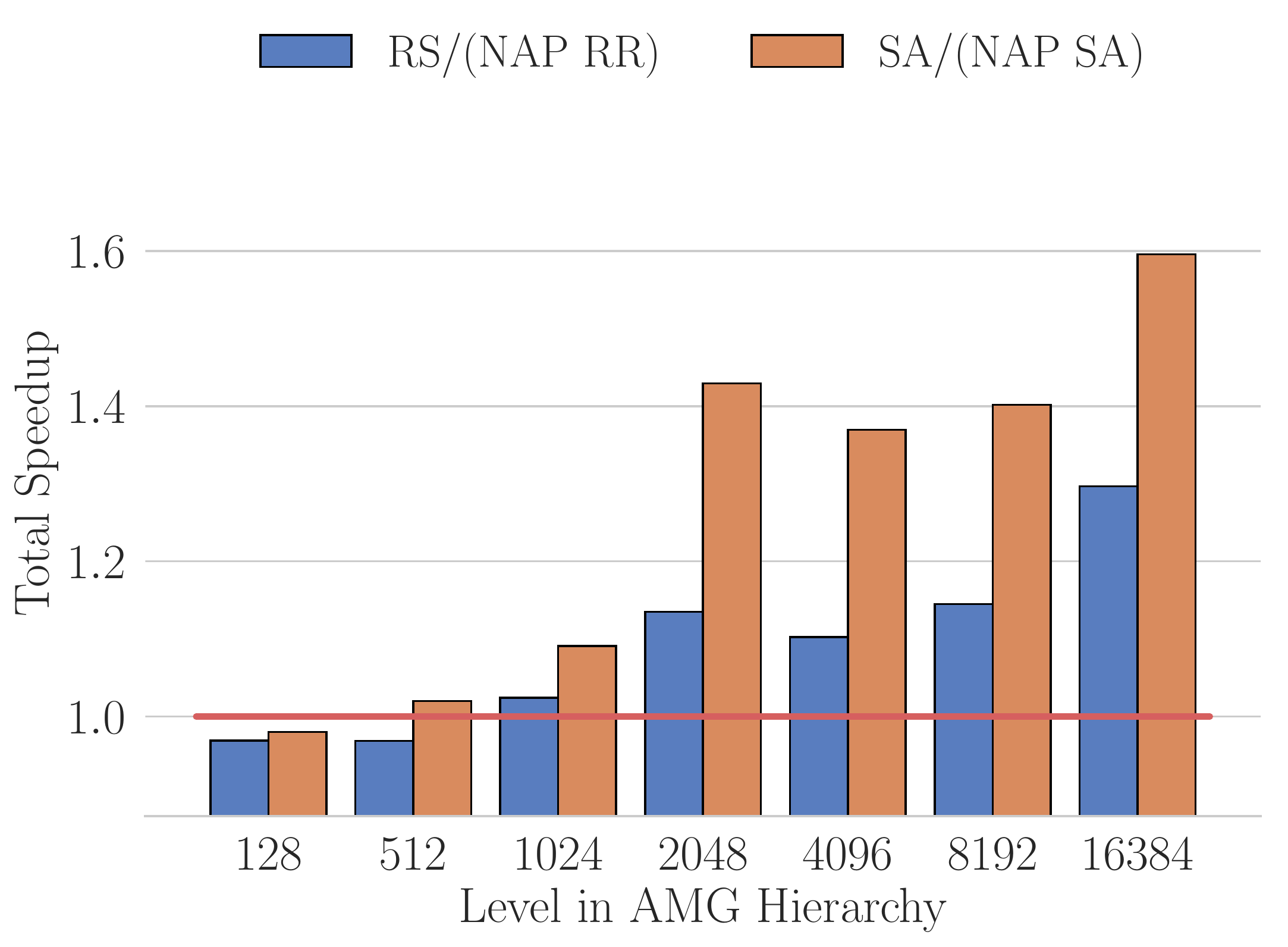}
        \caption{\texttt{Node-Aware Speedup}}
    \end{subfigure}
    \caption{Total AMG times for both the Ruge-St\"{u}ben and Smoothed
    Aggregation hierarchies for {\bf MFEM DPG Laplace} on Blue
    Waters.}\label{figure:mfem_dg_diffusion_total_time}
    \Description{Total DPG Diffusion AMG on Blue Waters Times.}
\end{figure*}
\begin{figure*}[ht!]
    \captionsetup[subfigure]{justification=centering}
    \centering
    \begin{subfigure}{0.49\textwidth}
        \centering
        \includegraphics[width=0.75\linewidth]{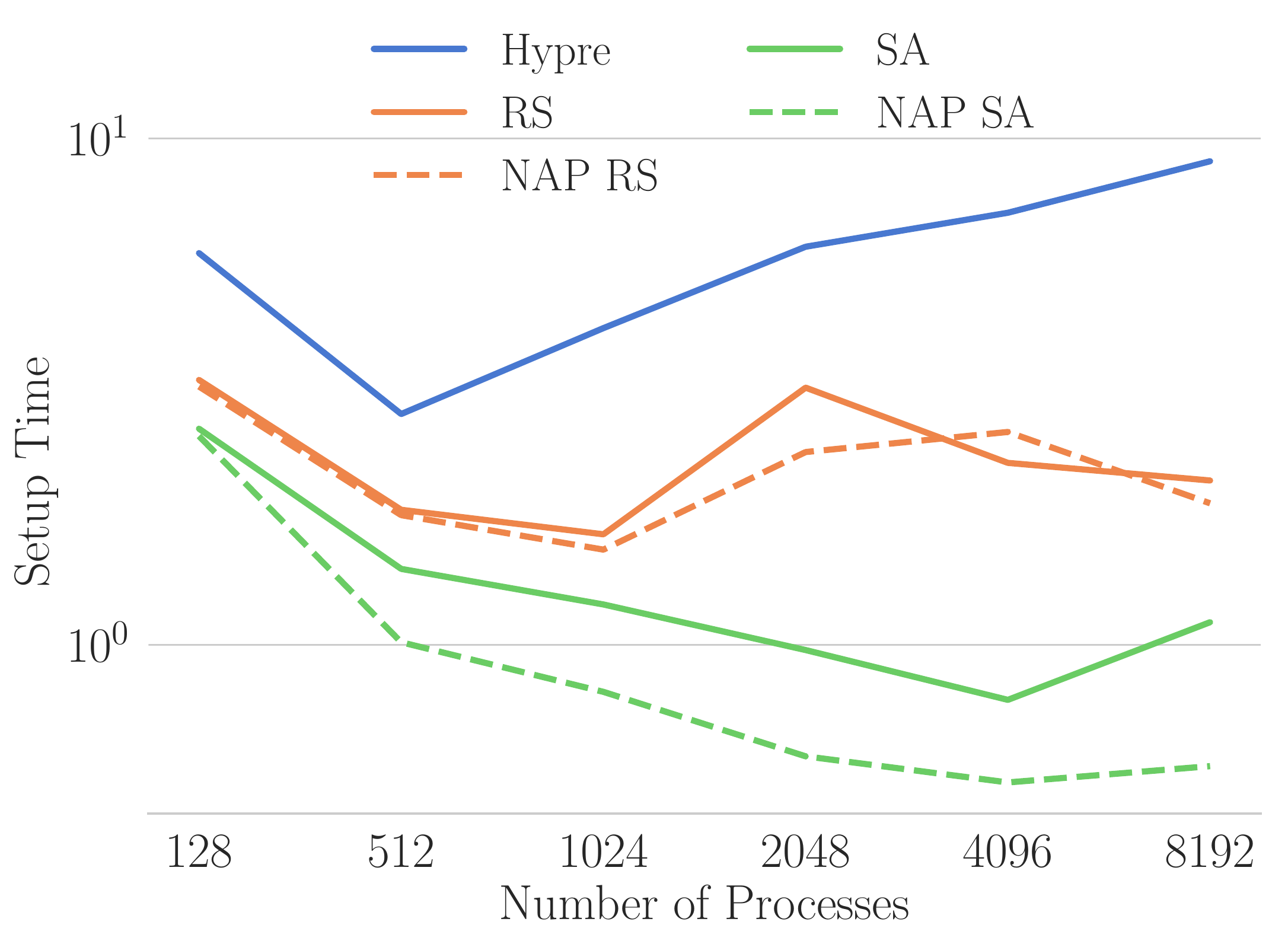}
        \caption{\texttt{Total Cost}}
    \end{subfigure}    \begin{subfigure}{0.49\textwidth}
        \centering
        \includegraphics[width=.75\linewidth]{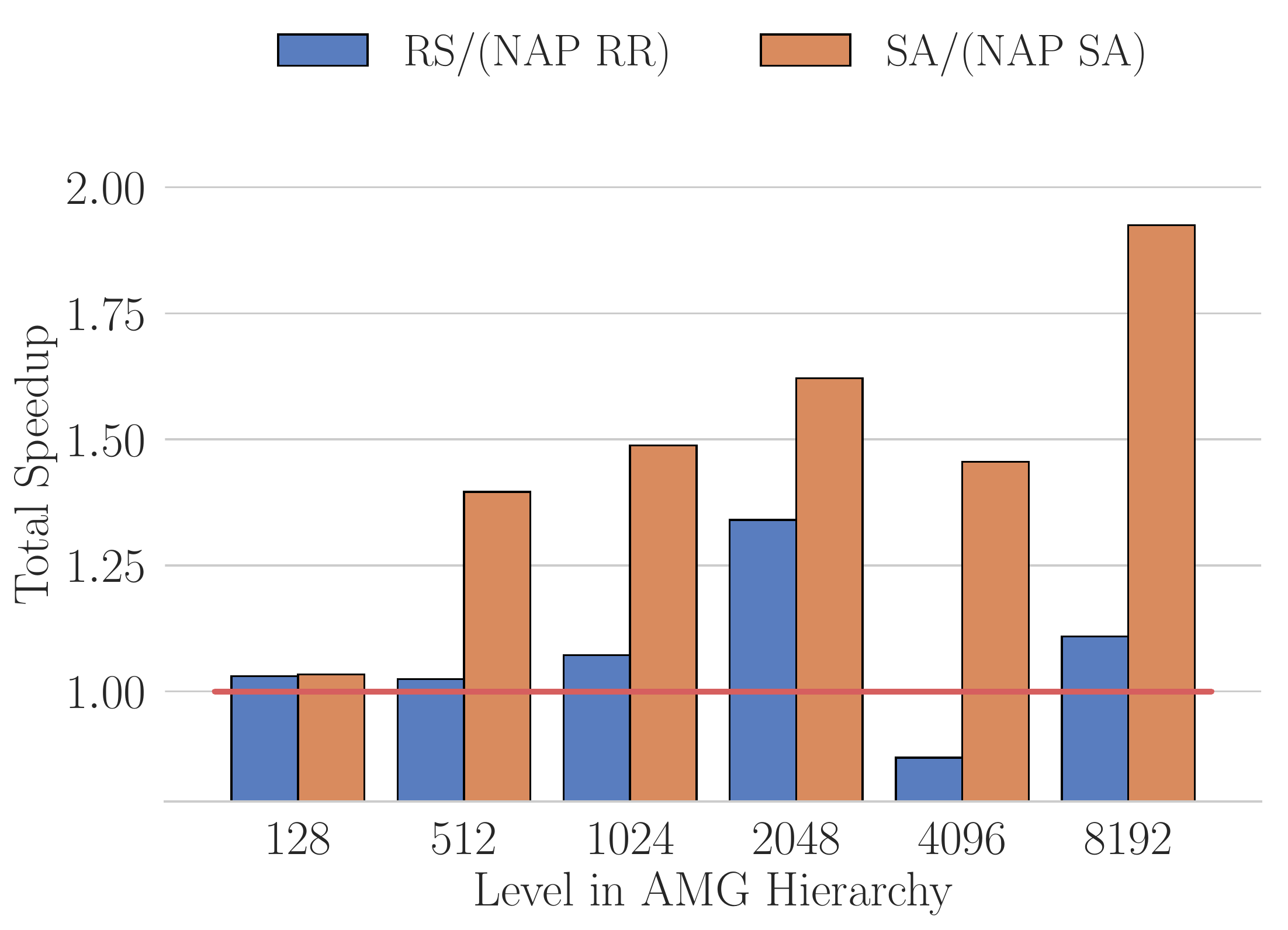}
        \caption{\texttt{Node-Aware Speedup}}
    \end{subfigure}
    \caption{Total AMG times for both the Ruge-St\"{u}ben and Smoothed
    Aggregation hierarchies for {\bf MFEM Grad-Div} on
    Quartz.}\label{figure:mfem_grad_div_quartz_total_time}
    \Description{Total Grad-Div AMG Times on Quartz.}
\end{figure*}

Node-aware communication can be used throughout the dominant methods of the
setup and solve phases.  This section presents performance and scaling results
of algebraic multigrid with node-aware communication.  Optimal strategies for
vector and matrix communication are determined during the formation of each
matrix in the AMG hierarchy.  After a matrix is created, the performance models
in Equations~\ref{eqn:standard_model},~\ref{eqn:NAP2_model},
and~\ref{eqn:NAP3_model} are calculated and the strategy with minimum modeled
cost is chosen.  Separate models are calculated for vector and matrix
communication, allowing for different strategies for each type of communication.

Throughout this section, both Ruge-St\"{u}ben and Smoothed Aggregation solvers
are analyzed for the problems that follow.
\begin{list}{}{\setlength{\leftmargin}{\parindent}\setlength{\itemindent}{-\parindent}}
\item {\bf MFEM Grad-Div -} The finite element discretization of $-\Grad(\alpha
    \Div(F)) + \beta F = f$, created with MFEM, on the three-dimensional
    fichera-q3 mesh.  The system has $2\,801\,664$ degrees-of-freedom and
    $117\,107\,712$ non-zeros, unless otherwise specified.
\item {\bf MFEM DPG Laplace -} The Discontinuous Petrov-Galerkin (DPG)
    discretization of the Laplace system $-\Laplace u = 1$, created with MFEM,
    on the three-dimensional star-q3 mesh.  This system contains $131\,720$ rows
    and $104\,529\,920$ non-zeros, unless otherwise specified.
\end{list}
The Ruge-St\"{u}ben hierarchies for these systems are aggressively coarsened
with HMIS and extended+i interpolation, while smoothed aggregation hierarchies
are created with aggregates based on an MIS-2 of the graph.  All tests are
performed with RAPtor~\cite{RAPtor} and compared against an identical
Ruge-Stuben hierarchy that is created and solved with a state-of-the-field
solver, Hypre's Boomer AMG~\cite{hypre, BoomerAMG}.  Performance tests are run
on both Blue Waters, a Cray supercomputer at the National Center for
Supercomputing Applications~\cite{BlueWaters,bw-in-vetter13}, and Quartz, an
Intel Xeon E5 machine at Lawrence Livermore National Laboratory.  All Blue
Waters tests are performed with $16$ processes per node, while Quartz timings
are acquired with $32$ processes per node.\@.

\begin{figure*}[ht!]
    \captionsetup[subfigure]{justification=centering}
    \centering
    \begin{subfigure}{0.49\textwidth}
        \centering
        \includegraphics[width=.75\linewidth]{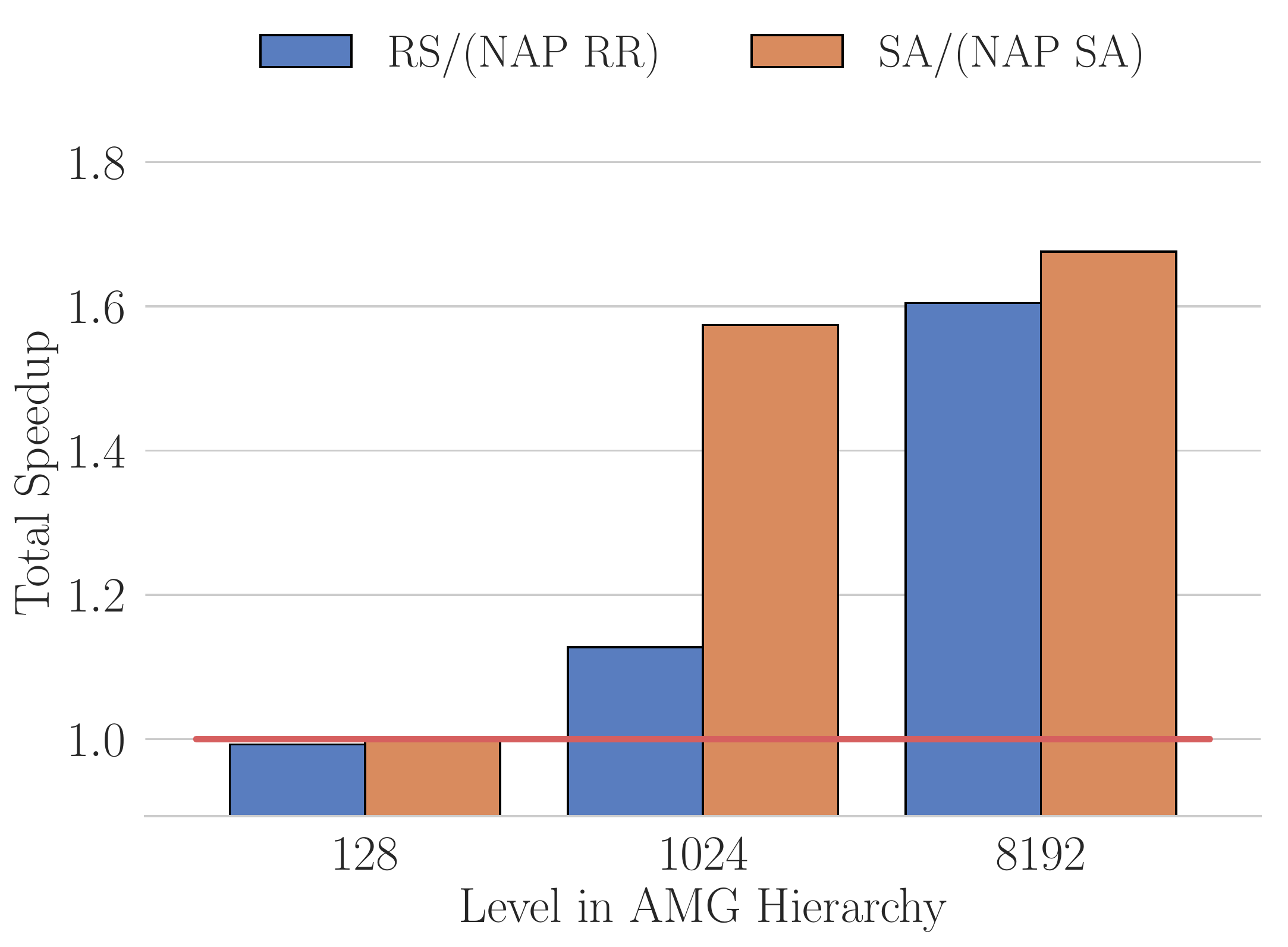}
        \caption{{\bf MFEM Grad-Div}}
    \end{subfigure}
    \begin{subfigure}{0.49\textwidth}
        \centering
        \includegraphics[width=.75\linewidth]{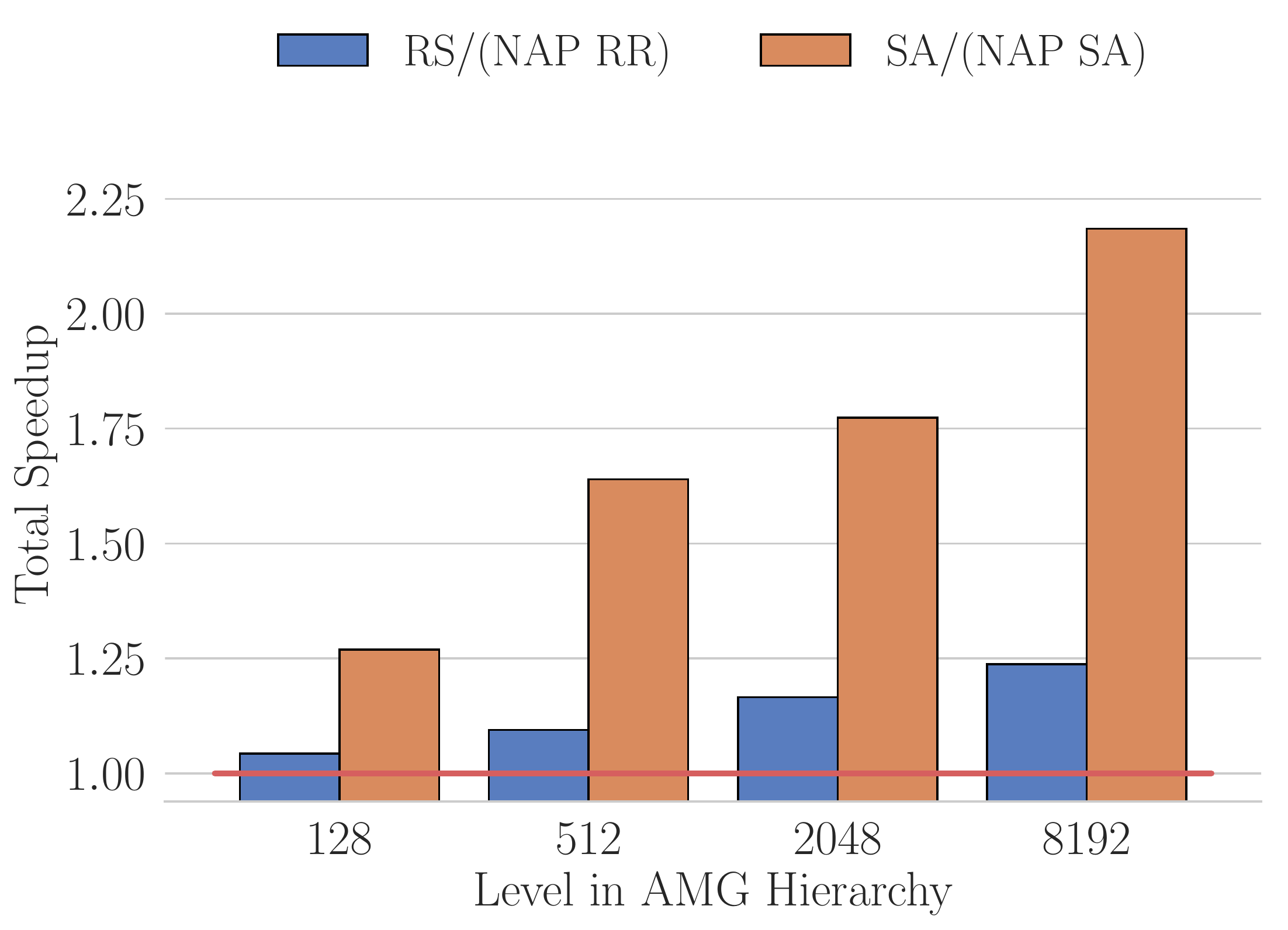}
        \caption{{\bf MFEM DPG Laplace}}
    \end{subfigure}
    \caption{Node-aware speedups achieved for both the Ruge-St\"{u}ben and Smoothed
    Aggregation hierarchies for weakly scaled MFEM systems.  All
    systems have approximately $10\,000$ degrees-of-freedom per
    core.}\label{figure:mfem_weak_total_time}
    \Description{Total AMG times, weakly scaled.}
\end{figure*}
\begin{figure*}[ht!]
    \captionsetup[subfigure]{justification=centering}
    \centering
    \begin{subfigure}{0.49\textwidth}
        \centering
        \includegraphics[width=0.75\linewidth]{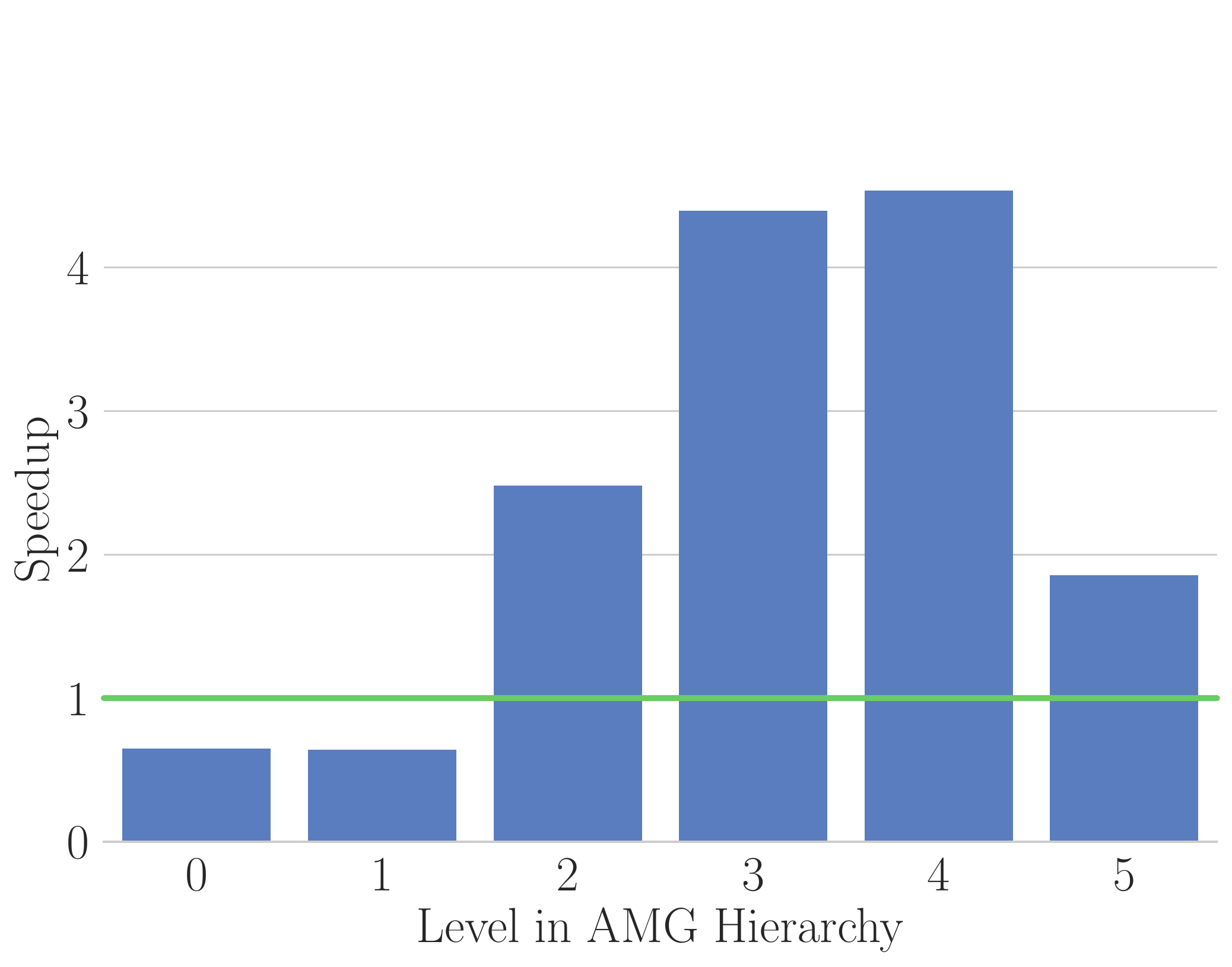}
        \caption{One Sweep}
    \end{subfigure}    \begin{subfigure}{0.49\textwidth}
        \centering
        \includegraphics[width=.75\linewidth]{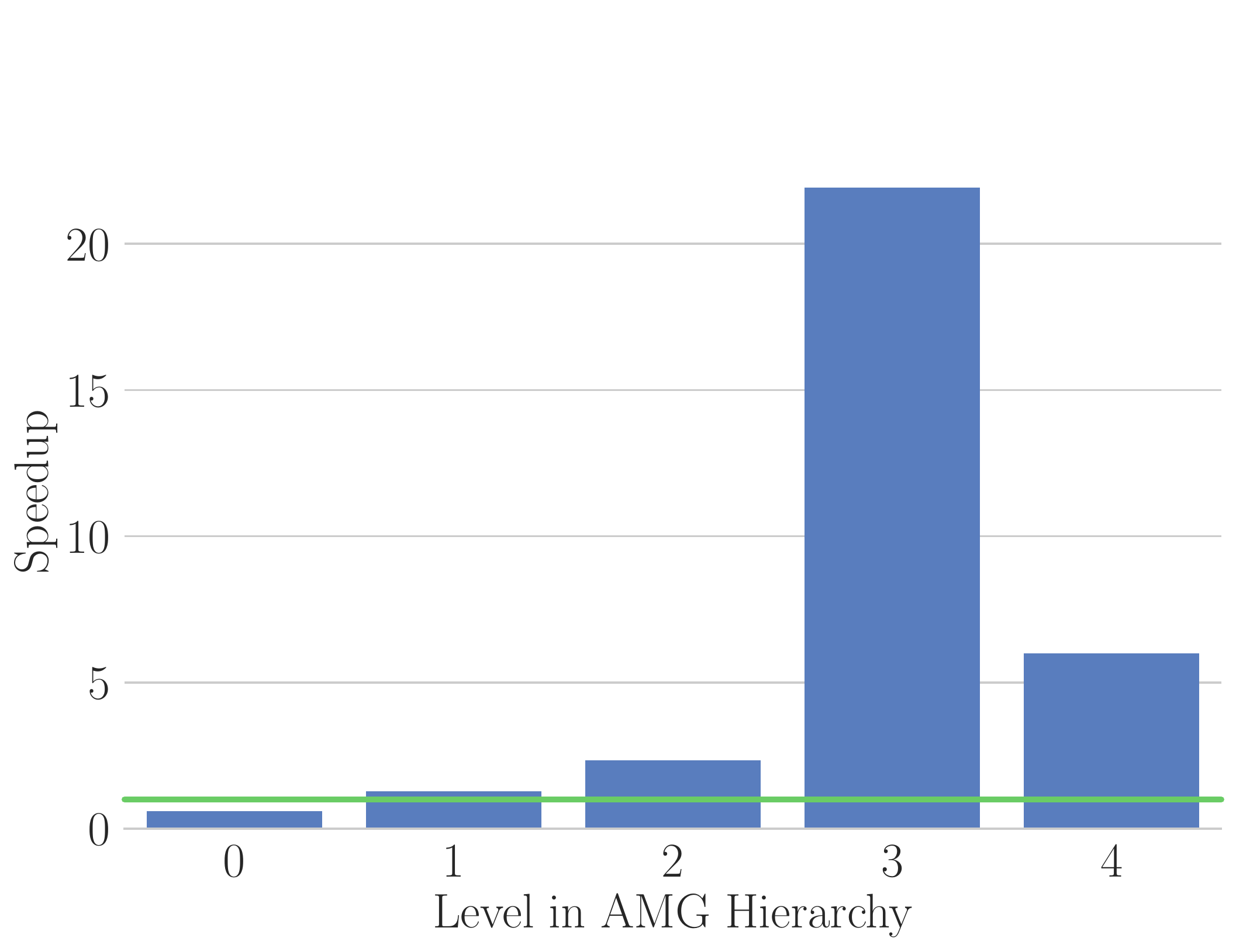}
        \caption{Two Sweeps}
    \end{subfigure}
    \caption{Speedup in $P^{T} \cdot (AP)$ throughout the smoothed aggregation
    hierarchy for a stenciled two-dimensional rotated anisotropic diffusion
    system when using a single sweep of Jacobi prolongation, as typical, or
    increasing to two smoothing sweeps.}\label{figure:multiple_sweeps}
    \Description{Galerkin Speedup.}
\end{figure*}

Figure~\ref{figure:mfem_laplace_setup_solve} displays the costs of both setting
up and solving Ruge-St\"{u}ben and smoothed aggregation hierarchies for the
{\bf MFEM Grad-Div} system on various core counts of Blue Waters.  Timings are
plotted both with and without node-aware communication.  While minimal
improvements are obtained at small core counts, node-aware communication yields
increased improvements as the problem is strongly scaled across processes,
particularly in the solve phase of AMG.  The cost of both determining the
appropriate communication strategy through models and forming node-aware
communicators is included in the setup phase costs.

The total cost of solving thh {\bf MFEM Grad-Div} system with AMG on {\bf Blue
Waters} are displayed in Figure~\ref{figure:mfem_grad_div_total_time} alongside
speedups achieved with node-aware communication.  Large improvements are
obtained over AMG with standard communication as the core count is increased,
with a nearly 4x speedup near the strong scaling limits of each solver.

The total AMG costs and node-aware speedups associated with solving the {\bf
MFEM DPG Laplace} system on {\bf Blue Waters} are displayed in
Figure~\ref{figure:mfem_dg_diffusion_total_time}.  While performance
improvements are less drastic than seen in the {\bf MFEM Grad-Div} system,
strong scalability is extended to at least $16\,384$ processes for both
solvers.  Furthermore, the speedups increase with process count.

The total AMG costs and node-aware speedups associated with solving the {\bf
MFEM Grad-Div} system on {\bf Quartz} are displayed in
Figure~\ref{figure:mfem_dg_diffusion_total_time}.  As with the Blue Waters
results, the total cost of AMG is improvement most drastically as the problem
is strongly scaled across the processors.

Similar performance is obtained when weakly scaling the system, as shown in
Figure~\ref{figure:mfem_weak_total_time}.  Node-aware communciation improves
performance at the various weak scales, with more drastic improvements as the
scale increases.  

Further improvements to node-aware AMG are possible, as many operations yield
little to no speedup, particularly in the setup phase.  While the majority of
operations throughout the setup phase are dependent on the sparsity pattern of
$A$, the transpose multiplication step $P^{T} \cdot AP$ is dependent on the
sparsity pattern of $P$.  Therefore, node-aware communication will have larger
improvements for transpose multiplication with denser $P$ matrices.  This
motivates increasing the density of $P$ to increase the accuracy of projecting
data between methods, such as using multiple sweeps of Jacobi
smoothing during the smoothed aggregation AMG setup.
Figure~\ref{figure:multiple_sweeps} shows the speedups obtained with
node-aware communication during $P^{T} \cdot AP$ on each level for both the
standard hierarchy and when using multiple smoothing sweeps.  There
is significant speedup for the denser $P$ resulting from the latter.

\section{Conclusions}\label{section:conclusion}

This paper has introduced a method of using node-awareness to reduce the amount
of inter-node communication at a trade-off of less costly intra-node
communication, and applied this communication method to the various parts of
algebraic multigrid.  Node-aware communication yields improvements in the
performance and scalability of both the setup and solve phases for a variety of three-dimensional matrices created with MFEM.  Node-aware communication yields improvements over the the state-of-the-field solver, Hypre.  Furthermore, performance and scalability improvements are obtained on both Blue Waters and Quartz. Future work will be done to further improve the performance of node-aware communication by improving the underlying performance models and allowing for different communication strategies for each node.  Furthermore, this strategy can be extended to a variety of other methods as well as emerging architectures.

\begin{acks}
This research is part of the Blue Waters sustained-petascale computing project,
which is supported by the National Science Foundation (awards OCI-0725070
and ACI-1238993) and the state of Illinois. Blue Waters is a joint effort of
the University of Illinois at Urbana-Champaign and its National Center for
Supercomputing Applications.  This material is based in part upon work
supported by the National Science Foundation Graduate Research Fellowship
Program under Grant Number DGE-1144245.  This material is based in part upon
work supported by the Department of Energy, National Nuclear Security
Administration, under Award Number DE-NA0002374. 
\end{acks}

\bibliographystyle{ACM-Reference-Format}
\bibliography{refs}


\begin{thebibliography}{33}


\ifx \showCODEN    \undefined \def \showCODEN     #1{\unskip}     \fi
\ifx \showDOI      \undefined \def \showDOI       #1{#1}\fi
\ifx \showISBNx    \undefined \def \showISBNx     #1{\unskip}     \fi
\ifx \showISBNxiii \undefined \def \showISBNxiii  #1{\unskip}     \fi
\ifx \showISSN     \undefined \def \showISSN      #1{\unskip}     \fi
\ifx \showLCCN     \undefined \def \showLCCN      #1{\unskip}     \fi
\ifx \shownote     \undefined \def \shownote      #1{#1}          \fi
\ifx \showarticletitle \undefined \def \showarticletitle #1{#1}   \fi
\ifx \showURL      \undefined \def \showURL       {\relax}        \fi
\providecommand\bibfield[2]{#2}
\providecommand\bibinfo[2]{#2}
\providecommand\natexlab[1]{#1}
\providecommand\showeprint[2][]{arXiv:#2}

\bibitem[\protect\citeauthoryear{??}{Bie}{[n. d.]}]%
        {Bienz_napspmv}
 \bibinfo{year}{[n. d.]}\natexlab{}.
\newblock  (\bibinfo{year}{[n. d.]}).
\newblock


\bibitem[\protect\citeauthoryear{??}{Blu}{[n. d.]}]%
        {BlueWaters}
 \bibinfo{year}{[n. d.]}\natexlab{}.
\newblock \bibinfo{title}{{Blue Waters}}.
\newblock \bibinfo{howpublished}{\url{https://bluewaters.ncsa.illinois.edu/}}.
\newblock


\bibitem[\protect\citeauthoryear{??}{hyp}{[n. d.]}]%
        {hypre}
 \bibinfo{year}{[n. d.]}\natexlab{}.
\newblock \bibinfo{title}{{HYPRE}: High performance preconditioners}.
\newblock \bibinfo{howpublished}{\url{http://www.llnl.gov/CASC/hypre/}}.
\newblock


\bibitem[\protect\citeauthoryear{??}{mfe}{[n. d.]}]%
        {mfem-library}
 \bibinfo{year}{[n. d.]}\natexlab{}.
\newblock \bibinfo{title}{{MFEM}: Modular finite element methods}.
\newblock \bibinfo{howpublished}{\url{mfem.org}}.
\newblock


\bibitem[\protect\citeauthoryear{Baker, Gamblin, Schulz, and Yang}{Baker
  et~al\mbox{.}}{2011}]%
        {AMGHPC}
\bibfield{author}{\bibinfo{person}{Allison~H. Baker}, \bibinfo{person}{Todd
  Gamblin}, \bibinfo{person}{Martin Schulz}, {and}
  \bibinfo{person}{Ulrike~Meier Yang}.} \bibinfo{year}{2011}\natexlab{}.
\newblock \showarticletitle{Challenges of Scaling Algebraic Multigrid Across
  Modern Multicore Architectures}. In \bibinfo{booktitle}{\emph{Proceedings of
  the 2011 IEEE International Parallel \& Distributed Processing Symposium}}
  \emph{(\bibinfo{series}{IPDPS '11})}. \bibinfo{publisher}{IEEE Computer
  Society}, \bibinfo{address}{Washington, DC, USA}, \bibinfo{pages}{275--286}.
\newblock
\showISBNx{978-0-7695-4385-7}
\urldef\tempurl%
\url{https://doi.org/10.1109/IPDPS.2011.35}
\showDOI{\tempurl}


\bibitem[\protect\citeauthoryear{Bhatele, Gamblin, Langer, Bremer, Draeger,
  Hamann, Isaacs, Landge, Levine, Pascucci, Schulz, and Still}{Bhatele
  et~al\mbox{.}}{2012}]%
        {SubcommTaskMap}
\bibfield{author}{\bibinfo{person}{Abhinav Bhatele}, \bibinfo{person}{Todd
  Gamblin}, \bibinfo{person}{Steven~H. Langer}, \bibinfo{person}{Peer-Timo
  Bremer}, \bibinfo{person}{Erik~W. Draeger}, \bibinfo{person}{Bernd Hamann},
  \bibinfo{person}{Katherine~E. Isaacs}, \bibinfo{person}{Aaditya~G. Landge},
  \bibinfo{person}{Joshua~A. Levine}, \bibinfo{person}{Valerio Pascucci},
  \bibinfo{person}{Martin Schulz}, {and} \bibinfo{person}{Charles~H. Still}.}
  \bibinfo{year}{2012}\natexlab{}.
\newblock \showarticletitle{Mapping Applications with Collectives over
  Sub-communicators on Torus Networks}. In
  \bibinfo{booktitle}{\emph{Proceedings of the International Conference on High
  Performance Computing, Networking, Storage and Analysis}}
  \emph{(\bibinfo{series}{SC '12})}. \bibinfo{publisher}{IEEE Computer Society
  Press}, \bibinfo{address}{Los Alamitos, CA, USA}, Article
  \bibinfo{articleno}{97}, \bibinfo{numpages}{11}~pages.
\newblock
\showISBNx{978-1-4673-0804-5}
\urldef\tempurl%
\url{http://dl.acm.org/citation.cfm?id=2388996.2389128}
\showURL{%
\tempurl}


\bibitem[\protect\citeauthoryear{Bhatele and Kale}{Bhatele and Kale}{2008}]%
        {TopoTaskMap}
\bibfield{author}{\bibinfo{person}{A. Bhatele} {and} \bibinfo{person}{L.~V.
  Kale}.} \bibinfo{year}{2008}\natexlab{}.
\newblock \showarticletitle{Application-specific topology-aware mapping for
  three dimensional topologies}. In \bibinfo{booktitle}{\emph{2008 IEEE
  International Symposium on Parallel and Distributed Processing}}.
  \bibinfo{pages}{1--8}.
\newblock
\showISSN{1530-2075}
\urldef\tempurl%
\url{https://doi.org/10.1109/IPDPS.2008.4536348}
\showDOI{\tempurl}


\bibitem[\protect\citeauthoryear{Bienz, Falgout, Gropp, Olson, and
  Schroder}{Bienz et~al\mbox{.}}{2016}]%
        {Bienz_sparsegal}
\bibfield{author}{\bibinfo{person}{Amanda Bienz}, \bibinfo{person}{Robert~D.
  Falgout}, \bibinfo{person}{William Gropp}, \bibinfo{person}{Luke~N. Olson},
  {and} \bibinfo{person}{Jacob~B. Schroder}.} \bibinfo{year}{2016}\natexlab{}.
\newblock \showarticletitle{Reducing Parallel Communication in Algebraic
  Multigrid through Sparsification}.
\newblock \bibinfo{journal}{\emph{SIAM Journal on Scientific Computing}}
  \bibinfo{volume}{38}, \bibinfo{number}{5} (\bibinfo{year}{2016}),
  \bibinfo{pages}{S332--S357}.
\newblock
\urldef\tempurl%
\url{https://doi.org/10.1137/15M1026341}
\showDOI{\tempurl}
\showeprint{https://doi.org/10.1137/15M1026341}


\bibitem[\protect\citeauthoryear{Bienz, Gropp, and Olson}{Bienz
  et~al\mbox{.}}{2018}]%
        {BienzEuroMPI}
\bibfield{author}{\bibinfo{person}{Amanda Bienz}, \bibinfo{person}{William~D.
  Gropp}, {and} \bibinfo{person}{Luke~N. Olson}.}
  \bibinfo{year}{2018}\natexlab{}.
\newblock \showarticletitle{Improving Performance Models for Irregular
  Point-to-Point Communication}. In \bibinfo{booktitle}{\emph{Proceedings of
  the 25th European {MPI} Users' Group Meeting, Barcelona, Spain, September
  23-26, 2018}}. \bibinfo{pages}{7:1--7:8}.
\newblock
\urldef\tempurl%
\url{https://doi.org/10.1145/3236367.3236368}
\showDOI{\tempurl}


\bibitem[\protect\citeauthoryear{Bienz and Olson}{Bienz and Olson}{2017}]%
        {RAPtor}
\bibfield{author}{\bibinfo{person}{Amanda Bienz} {and} \bibinfo{person}{Luke~N.
  Olson}.} \bibinfo{year}{2017}\natexlab{}.
\newblock \bibinfo{title}{{RAPtor}: parallel algebraic multigrid v0.1}.
\newblock
\newblock
\urldef\tempurl%
\url{https://github.com/lukeolson/raptor}
\showURL{%
\tempurl}
\newblock
\shownote{Release 0.1.}


\bibitem[\protect\citeauthoryear{Bode, Butler, Dunning, Hoefler, Kramer, Gropp,
  and Hwu}{Bode et~al\mbox{.}}{2013}]%
        {bw-in-vetter13}
\bibfield{author}{\bibinfo{person}{Brett Bode}, \bibinfo{person}{Michelle
  Butler}, \bibinfo{person}{Thom Dunning}, \bibinfo{person}{Torsten Hoefler},
  \bibinfo{person}{William Kramer}, \bibinfo{person}{William Gropp}, {and}
  \bibinfo{person}{Wen{-mei} Hwu}.} \bibinfo{year}{2013}\natexlab{}.
\newblock \showarticletitle{The {B}lue {W}aters Super-System for
  Super-Science}.
\newblock In \bibinfo{booktitle}{\emph{Contemporary High Performance Computing:
  From Petascale Toward Exascale} (\bibinfo{edition}{1} ed.)},
  \bibfield{editor}{\bibinfo{person}{Jeffrey~S. Vetter}} (Ed.).
  \bibinfo{series}{CRC Computational Science Series}, Vol.~\bibinfo{volume}{1}.
  \bibinfo{publisher}{Taylor and Francis}, \bibinfo{address}{Boca Raton},
  \bibinfo{pages}{339--366}.
\newblock
\urldef\tempurl%
\url{http://j.mp/RrBdPZ}
\showURL{%
\tempurl}


\bibitem[\protect\citeauthoryear{Brandt, McCormick, and Ruge}{Brandt
  et~al\mbox{.}}{1984}]%
        {BrMcRu1984}
\bibfield{author}{\bibinfo{person}{A. Brandt}, \bibinfo{person}{S.~F.
  McCormick}, {and} \bibinfo{person}{J.~W. Ruge}.}
  \bibinfo{year}{1984}\natexlab{}.
\newblock \showarticletitle{Algebraic Multigrid ({AMG}) for Sparse Matrix
  Equations}.
\newblock In \bibinfo{booktitle}{\emph{Sparsity and Its Applications}},
  \bibfield{editor}{\bibinfo{person}{D.~J. Evans}} (Ed.).
  \bibinfo{publisher}{Cambridge Univ. Press}, \bibinfo{address}{Cambridge},
  \bibinfo{pages}{257--284}.
\newblock


\bibitem[\protect\citeauthoryear{\c{C}ataly\"{u}rek and
  Aykanat}{\c{C}ataly\"{u}rek and Aykanat}{1999}]%
        {HypergraphPart}
\bibfield{author}{\bibinfo{person}{U.~V. \c{C}ataly\"{u}rek} {and}
  \bibinfo{person}{C. Aykanat}.} \bibinfo{year}{1999}\natexlab{}.
\newblock \showarticletitle{Hypergraph-partitioning-based decomposition for
  parallel sparse-matrix vector multiplication}.
\newblock \bibinfo{journal}{\emph{IEEE Transactions on Parallel and Distributed
  Systems}} \bibinfo{volume}{10}, \bibinfo{number}{7} (\bibinfo{date}{Jul}
  \bibinfo{year}{1999}), \bibinfo{pages}{673--693}.
\newblock
\showISSN{1045-9219}
\urldef\tempurl%
\url{https://doi.org/10.1109/71.780863}
\showDOI{\tempurl}


\bibitem[\protect\citeauthoryear{Falgout and Schroder}{Falgout and
  Schroder}{2014}]%
        {NonGal_Schroder}
\bibfield{author}{\bibinfo{person}{Robert~D. Falgout} {and}
  \bibinfo{person}{Jacob~B. Schroder}.} \bibinfo{year}{2014}\natexlab{}.
\newblock \showarticletitle{Non-{G}alerkin Coarse Grids for Algebraic
  Multigrid}.
\newblock \bibinfo{journal}{\emph{SIAM Journal on Scientific Computing}}
  \bibinfo{volume}{36}, \bibinfo{number}{3} (\bibinfo{year}{2014}),
  \bibinfo{pages}{C309--C334}.
\newblock
\urldef\tempurl%
\url{https://doi.org/10.1137/130931539}
\showDOI{\tempurl}
\showeprint{http://dx.doi.org/10.1137/130931539}


\bibitem[\protect\citeauthoryear{Gropp, Lusk, Doss, and Skjellum}{Gropp
  et~al\mbox{.}}{1996}]%
        {MPICH}
\bibfield{author}{\bibinfo{person}{William Gropp}, \bibinfo{person}{Ewing
  Lusk}, \bibinfo{person}{Nathan Doss}, {and} \bibinfo{person}{Anthony
  Skjellum}.} \bibinfo{year}{1996}\natexlab{}.
\newblock \showarticletitle{A high-performance, portable implementation of the
  MPI message passing interface standard}.
\newblock \bibinfo{journal}{\emph{Parallel Comput.}} \bibinfo{volume}{22},
  \bibinfo{number}{6} (\bibinfo{year}{1996}), \bibinfo{pages}{789 -- 828}.
\newblock
\showISSN{0167-8191}
\urldef\tempurl%
\url{https://doi.org/10.1016/0167-8191(96)00024-5}
\showDOI{\tempurl}


\bibitem[\protect\citeauthoryear{Gropp, Olson, and Samfass}{Gropp
  et~al\mbox{.}}{2016}]%
        {MaxRate}
\bibfield{author}{\bibinfo{person}{William Gropp}, \bibinfo{person}{Luke~N.
  Olson}, {and} \bibinfo{person}{Philipp Samfass}.}
  \bibinfo{year}{2016}\natexlab{}.
\newblock \showarticletitle{Modeling MPI Communication Performance on SMP
  Nodes: Is It Time to Retire the Ping Pong Test}. In
  \bibinfo{booktitle}{\emph{Proceedings of the 23rd European MPI Users' Group
  Meeting}} \emph{(\bibinfo{series}{EuroMPI 2016})}. \bibinfo{publisher}{ACM},
  \bibinfo{address}{New York, NY, USA}, \bibinfo{pages}{41--50}.
\newblock
\showISBNx{978-1-4503-4234-6}
\urldef\tempurl%
\url{https://doi.org/10.1145/2966884.2966919}
\showDOI{\tempurl}


\bibitem[\protect\citeauthoryear{Hendrickson and Kolda}{Hendrickson and
  Kolda}{2000}]%
        {Hendrickson}
\bibfield{author}{\bibinfo{person}{Bruce Hendrickson} {and}
  \bibinfo{person}{Tamara~G. Kolda}.} \bibinfo{year}{2000}\natexlab{}.
\newblock \showarticletitle{Graph Partitioning Models for Parallel Computing}.
\newblock \bibinfo{journal}{\emph{Parallel Comput.}} \bibinfo{volume}{26},
  \bibinfo{number}{12} (\bibinfo{date}{Nov} \bibinfo{year}{2000}),
  \bibinfo{pages}{1519--1534}.
\newblock
\showISSN{0167-8191}
\urldef\tempurl%
\url{https://doi.org/10.1016/S0167-8191(00)00048-X}
\showDOI{\tempurl}


\bibitem[\protect\citeauthoryear{Henson and Yang}{Henson and Yang}{2002}]%
        {BoomerAMG}
\bibfield{author}{\bibinfo{person}{Van~Emden Henson} {and}
  \bibinfo{person}{Ulrike~Meier Yang}.} \bibinfo{year}{2002}\natexlab{}.
\newblock \showarticletitle{{BoomerAMG}: A Parallel Algebraic Multigrid Solver
  and Preconditioner}.
\newblock \bibinfo{journal}{\emph{Appl. Numer. Math.}} \bibinfo{volume}{41},
  \bibinfo{number}{1} (\bibinfo{date}{April} \bibinfo{year}{2002}),
  \bibinfo{pages}{155--177}.
\newblock
\showISSN{0168-9274}
\urldef\tempurl%
\url{https://doi.org/10.1016/S0168-9274(01)00115-5}
\showDOI{\tempurl}


\bibitem[\protect\citeauthoryear{Karonis, de~Supinski, Foster, Gropp, Lusk, and
  Bresnahan}{Karonis et~al\mbox{.}}{2000}]%
        {Karonis}
\bibfield{author}{\bibinfo{person}{N.~T. Karonis}, \bibinfo{person}{B.~R. de
  Supinski}, \bibinfo{person}{I. Foster}, \bibinfo{person}{W. Gropp},
  \bibinfo{person}{E. Lusk}, {and} \bibinfo{person}{J. Bresnahan}.}
  \bibinfo{year}{2000}\natexlab{}.
\newblock \showarticletitle{Exploiting hierarchy in parallel computer networks
  to optimize collective operation performance}. In
  \bibinfo{booktitle}{\emph{Proceedings 14th International Parallel and
  Distributed Processing Symposium. IPDPS 2000}}. \bibinfo{pages}{377--384}.
\newblock
\urldef\tempurl%
\url{https://doi.org/10.1109/IPDPS.2000.846009}
\showDOI{\tempurl}


\bibitem[\protect\citeauthoryear{Kielmann, Hofman, Bal, Plaat, and
  Bhoedjang}{Kielmann et~al\mbox{.}}{1999}]%
        {Kielmann}
\bibfield{author}{\bibinfo{person}{Thilo Kielmann}, \bibinfo{person}{Rutger
  F.~H. Hofman}, \bibinfo{person}{Henri~E. Bal}, \bibinfo{person}{Aske Plaat},
  {and} \bibinfo{person}{Raoul A.~F. Bhoedjang}.}
  \bibinfo{year}{1999}\natexlab{}.
\newblock \showarticletitle{MagPIe: MPI's Collective Communication Operations
  for Clustered Wide Area Systems}.
\newblock \bibinfo{journal}{\emph{SIGPLAN Not.}} \bibinfo{volume}{34},
  \bibinfo{number}{8} (\bibinfo{date}{May} \bibinfo{year}{1999}),
  \bibinfo{pages}{131--140}.
\newblock
\showISSN{0362-1340}
\urldef\tempurl%
\url{https://doi.org/10.1145/329366.301116}
\showDOI{\tempurl}


\bibitem[\protect\citeauthoryear{McCormick and Ruge}{McCormick and
  Ruge}{1982}]%
        {McRu1982}
\bibfield{author}{\bibinfo{person}{S.~F. McCormick} {and}
  \bibinfo{person}{J.~W. Ruge}.} \bibinfo{year}{1982}\natexlab{}.
\newblock \showarticletitle{Multigrid methods for variational problems}.
\newblock \bibinfo{journal}{\emph{SIAM J. Numer. Anal.}} \bibinfo{volume}{19},
  \bibinfo{number}{5} (\bibinfo{year}{1982}), \bibinfo{pages}{924--929}.
\newblock


\bibitem[\protect\citeauthoryear{Pinar and Aykanat}{Pinar and Aykanat}{2004}]%
        {Pinar}
\bibfield{author}{\bibinfo{person}{Ali Pinar} {and} \bibinfo{person}{Cevdet
  Aykanat}.} \bibinfo{year}{2004}\natexlab{}.
\newblock \showarticletitle{Fast Optimal Load Balancing Algorithms for 1D
  Partitioning}.
\newblock \bibinfo{journal}{\emph{J. Parallel Distrib. Comput.}}
  \bibinfo{volume}{64}, \bibinfo{number}{8} (\bibinfo{date}{Aug.}
  \bibinfo{year}{2004}), \bibinfo{pages}{974--996}.
\newblock
\showISSN{0743-7315}
\urldef\tempurl%
\url{https://doi.org/10.1016/j.jpdc.2004.05.003}
\showDOI{\tempurl}


\bibitem[\protect\citeauthoryear{Ruge and St{\"{u}}ben}{Ruge and
  St{\"{u}}ben}{1987}]%
        {RuStu1987}
\bibfield{author}{\bibinfo{person}{J.~W. Ruge} {and} \bibinfo{person}{K.
  St{\"{u}}ben}.} \bibinfo{year}{1987}\natexlab{}.
\newblock \showarticletitle{Algebraic Multigrid ({AMG})}.
\newblock In \bibinfo{booktitle}{\emph{Multigrid Methods}},
  \bibfield{editor}{\bibinfo{person}{S.~F. McCormick}} (Ed.).
  \bibinfo{publisher}{SIAM}, \bibinfo{address}{Philadelphia},
  \bibinfo{pages}{73--130}.
\newblock


\bibitem[\protect\citeauthoryear{Sack and Gropp}{Sack and Gropp}{2012}]%
        {Sack}
\bibfield{author}{\bibinfo{person}{Paul Sack} {and} \bibinfo{person}{William
  Gropp}.} \bibinfo{year}{2012}\natexlab{}.
\newblock \showarticletitle{Faster Topology-aware Collective Algorithms Through
  Non-minimal Communication}. In \bibinfo{booktitle}{\emph{Proceedings of the
  17th ACM SIGPLAN Symposium on Principles and Practice of Parallel
  Programming}} \emph{(\bibinfo{series}{PPoPP '12})}. \bibinfo{publisher}{ACM},
  \bibinfo{address}{New York, NY, USA}, \bibinfo{pages}{45--54}.
\newblock
\showISBNx{978-1-4503-1160-1}
\urldef\tempurl%
\url{https://doi.org/10.1145/2145816.2145823}
\showDOI{\tempurl}


\bibitem[\protect\citeauthoryear{Solomonik, Bhatele, and Demmel}{Solomonik
  et~al\mbox{.}}{2011}]%
        {Solomonik}
\bibfield{author}{\bibinfo{person}{Edgar Solomonik}, \bibinfo{person}{Abhinav
  Bhatele}, {and} \bibinfo{person}{James Demmel}.}
  \bibinfo{year}{2011}\natexlab{}.
\newblock \showarticletitle{Improving Communication Performance in Dense Linear
  Algebra via Topology Aware Collectives}. In
  \bibinfo{booktitle}{\emph{Proceedings of 2011 International Conference for
  High Performance Computing, Networking, Storage and Analysis}}
  \emph{(\bibinfo{series}{SC '11})}. \bibinfo{publisher}{ACM},
  \bibinfo{address}{New York, NY, USA}, \bibinfo{pages}{77:1--77:11}.
\newblock
\showISBNx{978-1-4503-0771-0}
\urldef\tempurl%
\url{https://doi.org/10.1145/2063384.2063487}
\showDOI{\tempurl}


\bibitem[\protect\citeauthoryear{Sterck, Falgout, Nolting, and Yang}{Sterck
  et~al\mbox{.}}{2008}]%
        {DistTwo}
\bibfield{author}{\bibinfo{person}{Hans~De Sterck}, \bibinfo{person}{Robert~D.
  Falgout}, \bibinfo{person}{Joshua~W. Nolting}, {and}
  \bibinfo{person}{Ulrike~Meier Yang}.} \bibinfo{year}{2008}\natexlab{}.
\newblock \showarticletitle{Distance-two interpolation for parallel algebraic
  multigrid}.
\newblock \bibinfo{journal}{\emph{Numerical Linear Algebra with Applications}}
  \bibinfo{volume}{15}, \bibinfo{number}{2-3} (\bibinfo{year}{2008}),
  \bibinfo{pages}{115--139}.
\newblock
\showISSN{1099-1506}
\urldef\tempurl%
\url{https://doi.org/10.1002/nla.559}
\showDOI{\tempurl}


\bibitem[\protect\citeauthoryear{Sterck, Yang, and Heys}{Sterck
  et~al\mbox{.}}{2005}]%
        {AggCoarse2}
\bibfield{author}{\bibinfo{person}{Hans~De Sterck},
  \bibinfo{person}{Ulrike~Meier Yang}, {and} \bibinfo{person}{Jeffrey~J.
  Heys}.} \bibinfo{year}{2005}\natexlab{}.
\newblock \showarticletitle{Reducing Complexity in Parallel Algebraic Multigrid
  Preconditioners}.
\newblock \bibinfo{journal}{\emph{SIAM J. Matrix Anal. Appl.}}
  \bibinfo{volume}{27}, \bibinfo{number}{4} (\bibinfo{date}{Dec.}
  \bibinfo{year}{2005}), \bibinfo{pages}{1019--1039}.
\newblock
\showISSN{0895-4798}
\urldef\tempurl%
\url{https://doi.org/10.1137/040615729}
\showDOI{\tempurl}


\bibitem[\protect\citeauthoryear{Treister and Yavneh}{Treister and
  Yavneh}{2015}]%
        {NonGal_Treister}
\bibfield{author}{\bibinfo{person}{Eran Treister} {and} \bibinfo{person}{Irad
  Yavneh}.} \bibinfo{year}{2015}\natexlab{}.
\newblock \showarticletitle{Non-{G}alerkin Multigrid Based on Sparsified
  Smoothed Aggregation}.
\newblock \bibinfo{journal}{\emph{SIAM Journal on Scientific Computing}}
  \bibinfo{volume}{37}, \bibinfo{number}{1} (\bibinfo{year}{2015}),
  \bibinfo{pages}{A30--A54}.
\newblock
\urldef\tempurl%
\url{https://doi.org/10.1137/140952570}
\showDOI{\tempurl}
\showeprint{http://dx.doi.org/10.1137/140952570}


\bibitem[\protect\citeauthoryear{Tuminaro and Tong}{Tuminaro and Tong}{2000}]%
        {1592718}
\bibfield{author}{\bibinfo{person}{R.~S. Tuminaro} {and} \bibinfo{person}{C.
  Tong}.} \bibinfo{year}{2000}\natexlab{}.
\newblock \showarticletitle{Parallel Smoothed Aggregation Multigrid :
  Aggregation Strategies on Massively Parallel Machines}. In
  \bibinfo{booktitle}{\emph{Supercomputing, ACM/IEEE 2000 Conference}}.
  \bibinfo{pages}{5--5}.
\newblock
\showISSN{1063-9535}
\urldef\tempurl%
\url{https://doi.org/10.1109/SC.2000.10008}
\showDOI{\tempurl}


\bibitem[\protect\citeauthoryear{\"{U}m\.{i}t V.~\c{C}ataly\"{u}rek, Aykanat,
  and U\c{c}ar}{\"{U}m\.{i}t V.~\c{C}ataly\"{u}rek et~al\mbox{.}}{2010}]%
        {TwoDimPart}
\bibfield{author}{\bibinfo{person}{\"{U}m\.{i}t V.~\c{C}ataly\"{u}rek},
  \bibinfo{person}{Cevdet Aykanat}, {and} \bibinfo{person}{Bora U\c{c}ar}.}
  \bibinfo{year}{2010}\natexlab{}.
\newblock \showarticletitle{On Two-Dimensional Sparse Matrix Partitioning:
  Models, Methods, and a Recipe}.
\newblock \bibinfo{journal}{\emph{SIAM Journal on Scientific Computing}}
  \bibinfo{volume}{32}, \bibinfo{number}{2} (\bibinfo{year}{2010}),
  \bibinfo{pages}{656--683}.
\newblock
\urldef\tempurl%
\url{https://doi.org/10.1137/080737770}
\showDOI{\tempurl}
\showeprint{https://doi.org/10.1137/080737770}


\bibitem[\protect\citeauthoryear{Vastenhouw and Bisseling}{Vastenhouw and
  Bisseling}{2005}]%
        {BisselingDataDist}
\bibfield{author}{\bibinfo{person}{Brendan Vastenhouw} {and}
  \bibinfo{person}{Rob~H. Bisseling}.} \bibinfo{year}{2005}\natexlab{}.
\newblock \showarticletitle{A Two-Dimensional Data Distribution Method for
  Parallel Sparse Matrix-Vector Multiplication}.
\newblock \bibinfo{journal}{\emph{SIAM Rev.}} \bibinfo{volume}{47},
  \bibinfo{number}{1} (\bibinfo{year}{2005}), \bibinfo{pages}{67--95}.
\newblock
\urldef\tempurl%
\url{https://doi.org/10.1137/S0036144502409019}
\showDOI{\tempurl}
\showeprint{https://doi.org/10.1137/S0036144502409019}


\bibitem[\protect\citeauthoryear{Wesolowski, Venkataraman, Gupta, Yeom, Bisset,
  Sun, Jetley, Quinn, and Kale}{Wesolowski et~al\mbox{.}}{2014}]%
        {Tram}
\bibfield{author}{\bibinfo{person}{L. Wesolowski}, \bibinfo{person}{R.
  Venkataraman}, \bibinfo{person}{A. Gupta}, \bibinfo{person}{J.~S. Yeom},
  \bibinfo{person}{K. Bisset}, \bibinfo{person}{Y. Sun}, \bibinfo{person}{P.
  Jetley}, \bibinfo{person}{T.~R. Quinn}, {and} \bibinfo{person}{L.~V. Kale}.}
  \bibinfo{year}{2014}\natexlab{}.
\newblock \showarticletitle{TRAM: Optimizing Fine-Grained Communication with
  Topological Routing and Aggregation of Messages}. In
  \bibinfo{booktitle}{\emph{2014 43rd International Conference on Parallel
  Processing}}. \bibinfo{pages}{211--220}.
\newblock
\showISSN{0190-3918}
\urldef\tempurl%
\url{https://doi.org/10.1109/ICPP.2014.30}
\showDOI{\tempurl}


\bibitem[\protect\citeauthoryear{Yang}{Yang}{2010}]%
        {AggCoarse}
\bibfield{author}{\bibinfo{person}{Ulrike~Meier Yang}.}
  \bibinfo{year}{2010}\natexlab{}.
\newblock \showarticletitle{On long-range interpolation operators for
  aggressive coarsening}.
\newblock \bibinfo{journal}{\emph{Numerical Linear Algebra with Applications}}
  \bibinfo{volume}{17}, \bibinfo{number}{2--3} (\bibinfo{year}{2010}),
  \bibinfo{pages}{453--472}.
\newblock
\urldef\tempurl%
\url{https://doi.org/10.1002/nla.689}
\showDOI{\tempurl}


\end{thebibliography}

\end{document}